\documentclass[prd,amsfonts,onecolumn,superscriptaddress,aps,nofootinbib,11pt]{revtex4-1}

\usepackage{booktabs}
\usepackage{amsmath, amsthm, amssymb, amsfonts, mathrsfs}
\usepackage{physics, braket, slashed, siunitx, bm}
\usepackage{bbold}
\usepackage{listings}
\usepackage{color}
\usepackage[usenames,dvipsnames]{xcolor}
\usepackage{graphicx}
\usepackage{float}
\usepackage{placeins}
\usepackage{booktabs}
\usepackage{hyperref}
\usepackage{soul}
\usepackage{ulem}

\hypersetup{
    unicode=false,          
    pdftoolbar=true,        
    pdfmenubar=true,        
    pdffitwindow=false,     
    pdfauthor={William},     
    colorlinks=true,       
    linkcolor=blue,          
    citecolor=red,        
    urlcolor=blue
}

\providecommand{\be}{ \begin{equation} }
\providecommand{\ee}{\end{equation}}
\providecommand{\bea}{\begin{eqnarray}}
\providecommand{\eea}{\end{eqnarray}}
\providecommand{\nn}{\nonumber}

\def\331{SU(3)_C\otimes SU(3)_L\otimes U(1)_X}

\begin{document}

\title{Axion-Neutrino Interplay in a Gauged Two-Higgs-Doublet Model}
\author{Alex G. Dias}
\email{alex.dias@ufabc.edu.br}
\affiliation{Centro de Ci\^encias Naturais e Humanas, Universidade Federal do ABC,\\
09210-580, Santo Andr\'e-SP, Brasil}
\author{Julio Leite}
\email{julio.leite@ufabc.edu.br}
\affiliation{Centro de Ci\^encias Naturais e Humanas, Universidade Federal do ABC,\\
09210-580, Santo Andr\'e-SP, Brasil}
\author{Diego S. V. Gon\c{c}alves}
\email{diego.vieira@ufabc.edu.br}
\affiliation{Centro de Ci\^encias Naturais e Humanas, Universidade Federal do ABC,\\
09210-580, Santo Andr\'e-SP, Brasil}
\date{\today}

\setstcolor{magenta}

\begin{abstract}

We propose a gauged two-Higgs-doublet model (2HDM) featuring an anomalous Peccei-Quinn symmetry, $U(1)_{PQ}$. Dangerous tree-level flavour-changing neutral currents, common in 2HDMs, are forbidden by the extra gauge symmetry, $U(1)_X$. In our construction, the solutions to the important issues of neutrino masses, dark matter and the strong CP problem are interrelated. Neutrino masses are generated via a Dirac seesaw mechanism and are suppressed by the ratio of the $U(1)_X$ and the $U(1)_{PQ}$ breaking scales. Naturally small neutrino masses suggest that the breaking of $U(1)_X$ occurs at a relatively low scale, which may lead to observable signals in near-future experiments. Interestingly, spontaneous symmetry breaking does not lead to mixing between the $U(1)_X$ gauge boson, $Z^\prime$, and the standard $Z$. For the expected large values of the $U(1)_{PQ}$ scale, the associated axion becomes ``invisible'', with DFSZ-like couplings, and may account for the observed abundance of cold dark matter. Moreover, a viable parameter space region, which falls within the expected sensitivities of forthcoming axion searches, is identified. We also observe that the flavour-violating process of kaon decaying into pion plus axion, $K^+ \to \pi^+ a$, is further suppressed by the $U(1)_X$ scale, providing a rather weak lower bound for the axion decay constant $f_a$.

\end{abstract}

\maketitle

\section{Introduction}

The origin of small neutrino masses and the nature of dark matter (DM) are two of the most pressing issues with no answers within the Standard Model (SM) of particle physics. Nevertheless, there exist plenty of other open questions suggesting the need of physics beyond the SM, {\it e.g.} the non-observation of a CP-violating phase in the strong interaction sector.

The observation of neutrino oscillations \cite{Fukuda:1998mi,Ahmad:2001an,Ahmad:2002jz} has led to an understanding that, contrary to the SM picture, neutrinos are massive, albeit extremely light. A plethora of new physics proposals has been put forward to explain the smallness of neutrino masses: from the seesaw mechanism and its various realisations, see {\it e.g.} \cite{Valle:2015pba} -- relying on new physics at very large scales -- to radiative mechanisms \cite{Zee:1985id, Babu:1988ki, Pilaftsis:1991ug, Ma:2006km} -- taking place at much lower scales, possibly within experimental reach. On the experimental side, many efforts have been helping us to determine not only neutrino masses per se but also other intrinsically related properties, such as neutrino mass ordering and absolute scale, CP phase and whether neutrinos are their own anti-particles \cite{deSalas:2020pgw}. 

Another major drawback of the SM is the absence of a suitable candidate to account for the observed dark matter relic abundance \cite{Aghanim:2018eyx}, whose evidence arises from many sources \cite{Bertone:2016nfn}: from studies of galaxy rotation curves to cosmic microwave background data. Among the most appealing DM candidates, there are the weakly interacting massive particles or WIMPs, which, despite various experimental searches, have not yet been observed \cite{Aprile:2018dbl}. On the other hand, axions -- originally proposed as a key ingredient of the Peccei-Quinn (PQ) solution to the strong CP problem \cite{Peccei:1977hh, Weinberg:1977ma, Wilczek:1977pj} (for reviews, see \cite{Kim:2008hd,Hook:2018dlk}) -- define another well-known class of DM candidates \cite{Abbott:1982af,Preskill:1982cy,Dine:1982ah}, having the advantage of being capable to evade strong constraints coming from WIMP searches.

The exciting possibility of linking neutrino masses to dark matter and the strong CP problem via axions has been investigated in different scenarios.  For Majorana neutrinos, the implementation of seesaw mechanisms, where the large seesaw scale is identified with the PQ scale -- already considered many decades ago \cite{Langacker:1986rj, Mohapatra:1982tc} -- has been explored in various proposals more recently, see {\it e.g.} \cite{Dias:2014osa,Bertolini:2014aia,Clarke:2015bea,Ahn:2015pia, Ballesteros:2016euj,Ballesteros:2016xej}. Additionally, the case for Dirac neutrinos has also become the subject of several studies \cite{Chen:2012baa, Gu:2016hxh, Peinado:2019mrn, Baek:2019wdn, Dias:2020kbj}. The latter case has been attracting more attention over the last few years since experiments, such as searches for neutrinoless double beta decays \cite{KamLAND-Zen:2016pfg}, have not so far found any evidence for lepton number violation, which could confirm the Majorana nature of neutrinos.

In this work, we propose a model in which the issues of neutrino masses, dark matter and strong CP problem are addressed simultaneously. The SM group is enlarged by an extra gauge symmetry, $U(1)_X$, as well as the global Peccei-Quinn symmetry, $U(1)_{PQ}$. The model contains two Higgs doublets, as in two-Higgs-doublet models (2HDMs) \cite{Branco:2011iw}, plus two singlets. As a result of the $U(1)_X$ charge distribution, the Higgs doublets couple to different fermions, preventing the emergence of dangerous flavour-changing neutral currents (FCNCs) at tree level.  In the fermion sector, a minimal field content, including extra quarks and neutral leptons, is added to ensure gauge anomaly cancellation as well as a consistent generation of neutrino masses. Different constructions of 2HDMs with a $U(1)_X$ symmetry have already been proposed to explain the suppression of FCNCs \cite{Ko:2012hd}, along with the implementation of WIMP  \cite{Ko:2014uka} and axion \cite{Okada:2020cvq} dark matter candidates, seesaw mechanism for the neutrino masses \cite{Campos:2017dgc,Camargo:2018uzw,Cogollo:2019mbd} and other phenomenological issues \cite{Ko:2013zsa}.

Neutrino masses are generated via a Dirac seesaw mechanism, coming with the suppression factor $v_\varphi/v_\sigma$ , where $v_\varphi$ and $v_\sigma$ are the $U(1)_X$ and $U(1)_{PQ}$ scales, respectively. Naturally small neutrino masses suggest that the breaking of $U(1)_X$ occurs at a relatively low scale, $v_\varphi\ll v_\sigma$, which may lead to observable signals in near-future experiments. Interestingly, spontaneous symmetry breaking does not lead to mixing between the $U(1)_X$ gauge boson, $Z^\prime$, and the SM $Z$. For the expected large values of the $U(1)_{PQ}$ scale, $v_\sigma$, the associated axion becomes ``invisible'', with Dine-Fischler-Srednicki-Zhitnitsky (DFSZ)-like couplings \cite{Dine:1981rt, Zhitnitsky:1980tq}, and may account for the observed abundance of cold dark matter. Moreover, a viable parameter space region, which falls within the expected sensitivities of forthcoming axion searches, is identified. Therefore, the alluring axion-neutrino interplay renders the model theoretically consistent and phenomenologically rich.

The remaining of the paper is organised as follows. In Sec. \ref{sec:modbuild}, we discuss the model building rationale and present the field content and symmetry properties. In Sec. \ref{sec:scalar}, the scalar spectrum is derived, and the orthonormalisation of the Goldstone bosons is thoroughly discussed. Next, in Sec. \ref{sec:gaugespec}, we obtain the gauge spectrum, augmented by the new boson $Z^\prime$, and elucidate the non-mixing property between $Z$ and $Z^\prime$. The fermion sector is explored in Section \ref{sec:fermion}, where the mixing patterns between the extra and the standard fermions are obtained via the diagonalisation of their mass matrices. Naturally small neutrino masses are shown to be generated via a Dirac seesaw mechanism. We turn our attention to the axion physics in Sec. \ref{sec:axion}, where we show the main axion properties, derive the model-dependent axion couplings to photons and fermions as well as investigate relevant phenomenological consequences. Our final remarks are made in Sec. \ref{sec:conc}.

\section{Model building}\label{sec:modbuild}

We start by considering a 2HDM with a $U(1)_X$ gauge symmetry under which the Higgs doublets, $\Phi_u$ and $\Phi_d$, carry different charges, forbidding the appearance of dangerous Higgs-mediated FCNCs.
In addition to the Higgs doublets, we introduce a $SU(2)_L$ Higgs singlet, $\varphi$, also charged under the new local group.
We assume that $\varphi$ acquires a vacuum expectation value (vev) above the electroweak scale, spontaneously breaking $U(1)_X$ and generating a mass to the associated gauge boson $Z^\prime$.
Taking advantage of the presence of two Higgs doublets, fundamental ingredients of DFSZ-type axion models, an anomalous Peccei-Quinn (PQ) symmetry $U(1)_{PQ}$ is implemented. 
This is made possible with the introduction of a second singlet, $\sigma$, whose large vev breaks $U(1)_{PQ}$, triggering the PQ mechanism that solves the strong CP problem.
The (pseudo-) Goldstone boson of the $U(1)_{PQ}$ spontaneous breaking is identified with the axion field, and it can play the role of cold dark matter.

In order for the Peccei-Quinn symmetry to be realised {\it à la} DFSZ, the Higgs doublets must also be charged under $U(1)_{PQ}$, and each of them, namely $\Phi_u$ and $\Phi_d$, should couple to either the right-handed up-type quarks, $u_{aR}$, or down-type quarks, $d_{aR}$, respectively. However, if only standard quarks are present, $U(1)_X$ anomaly cancellation -- in particular, for the $[SU(3)_C]^2\times U(1)_X$ anomaly -- is achieved once the scalar doublets are identically charged under $U(1)_X$: $X_{\Phi_u}=X_{\Phi_d}$. Therefore, to have $X_{\Phi_u}\neq X_{\Phi_d}$ so that $U(1)_X$ is responsible for the absence of Higgs-mediated FCNCs, we need to extend the quark sector of our model in such a way that all $U(1)_X$ anomalies vanish. 

In the quest for minimal solutions, we add $n$ pairs of quarks (vector-like under the SM group), $k_{nL,R}$, carrying the same electric charge $q_k$, and try to find minimal sets of $(n, q_k)$ for which all $U(1)_X$ anomalies are cancelled. This is obviously only possible if the new quarks are chirally charged under $U(1)_X$, and, for simplicity, we assume that they get their $U(1)_X$ charges, as well as masses, via tree-level couplings to $\varphi$. We can divide our search into two major cases depending on whether the right-handed charged leptons, $e_{aR}$, couple to $\Phi_d$, as in the type-I DFSZ model or type-II 2HDM, or $\Phi_u$, as in the type-II DFSZ model or flipped (type-Y) 2HDM. In the former case, one of the simplest solutions is $(n, q_k)=(3, 2/3)$, whereas for the latter case, we find $(n, q_k)=(3, -1/3)$. In this work, we focus on the second case, which requires $n=3$ pairs of extra quarks carrying the same electric charge as the down-type quarks: $q_k=-1/3$. 

At last, we include three right-handed neutrinos, $\nu_{aR}$, and three pairs of neutral leptons $n_{aL,R}$, which are vector-like under the gauge symmetries. The presence of such fields allows for a consistent generation of small neutrino masses via a Dirac seesaw mechanism, taking place via an interplay among all scales in the model.

\begin{table}[ht]
\begin{tabular}{|c|c|c|c|c|}
\hline
          & $\mathcal{G}_{SM}$ & $U(1)_{global}$ & $U(1)_{afree}$                                \\\hline\hline
\,\,$L_{aL}$\,\,     & $({\bf 1},{\bf 2}, -1/2)$ & \,\,$q^\prime_{n_L}-q^\prime_{\Phi_u}-2q^\prime_{\sigma}-q^\prime_{\varphi}$ \,\,   & $-3l_{Q_L}$     \\\hline
$e_{aR}$  & $({\bf 1},{\bf 1}, -1)$     & \,\,$q^\prime_{n_L}-2q^\prime_{\Phi_u}-2q^\prime_{\sigma}-q^\prime_{\varphi}$\,\,  & $-3l_{Q_L}-l_{\Phi_u}$    \\\hline
$\nu_{aR}$  & $({\bf 1},{\bf 1}, 0)$    & $q^\prime_{n_L}-q^\prime_\varphi$  & $-3l_{Q_L}+l_{\Phi_u}$          \\\hline
$n_{aL}$  & $({\bf 1},{\bf 1}, 0)$      & $q^\prime_{n_L}$   & \,\,$-3l_{Q_L}+l_{\Phi_u}+l_\varphi$\,\,      \\\hline
$n_{aR}$  & $({\bf 1},{\bf 1}, 0)$      & $q^\prime_{n_L}-q^\prime_\sigma$ & $-3l_{Q_L}+l_{\Phi_u}+l_\varphi$                \\\hline\hline
$Q_{aL}$  & $({\bf 3},{\bf 2}, 1/6)$    & $q^\prime_{Q_L}$      & $l_{Q_L}$                                 \\\hline
$u_{aR}$  & $({\bf 3},{\bf 1}, 2/3)$    & $q^\prime_{Q_L}+q^\prime_{\Phi_u}$ & $l_{Q_L}+l_{\Phi_u}$ \\\hline
$d_{aR}$  & $({\bf 3},{\bf 1}, -1/3)$   & $q^\prime_{Q_L}-q^\prime_{\Phi_u}-q^\prime_{\sigma}-q^\prime_{\varphi}$ & $l_{Q_L}-l_{\Phi_u}-l_\varphi$   \\\hline
$k_{aL}$  & $({\bf 3},{\bf 1}, -1/3)$    & $q^\prime_{Q_L}-q^\prime_{\Phi_u}-q^\prime_{\sigma}-q^\prime_{\varphi}$        & \,\,$l_{Q_L}-l_{\Phi_u}-l_\varphi$\,\,         \\\hline
$k_{aR}$  & \,\,$({\bf 3},{\bf 1}, -1/3)$\,\,  & $q^\prime_{Q_L}-q^\prime_{\Phi_u}-q^\prime_{\sigma}$  & $l_{Q_L}-l_{\Phi_u}$                 \\\hline\hline
$\Phi_u$  & $({\bf 1},{\bf 2}, 1/2)$    & $q^\prime_{\Phi_u}$    & $l_{\Phi_u}$  \\\hline
$\Phi_d$  & $({\bf 1},{\bf 2}, 1/2)$    & $q^\prime_{\Phi_u}+ q^\prime_{\sigma}+q^\prime_\varphi$  & $l_{\Phi_u}+l_{\varphi}$     \\ \hline
$\varphi$ & $({\bf 1},{\bf 1}, 0)$      & $q^\prime_{\varphi}$   & $l_{\varphi}$                  \\\hline
$\sigma$  & $({\bf 1},{\bf 1}, 0)$      & $q^\prime_{\sigma}$    & $0$                            \\\hline
\end{tabular}
\caption{Fermions, scalars and their symmetry transformations. The $\mathcal{G}_{SM}$ column shows the field transformations under the SM group. In the $U(1)_{global}$ column, we present the independent charges associated with the five global symmetries of the model, including $U(1)_{PQ}$ for which $q^\prime_\sigma \equiv PQ_\sigma  \neq 0$. Amongst them, three satisfy anomaly-free conditions, displayed in the $U(1)_{afree}$ column, including $U(1)_{X}$ for which $l_\varphi \equiv X_\varphi  \neq 0$.}
\label{tab1}
\end{table}
In Table \ref{tab1}, we present the fermion and scalar contents together with their symmetry transformations. In the $U(1)_{global}$ column, we have five independent charges which can be linked to the five Abelian symmetries in the model: $U(1)_Y$, $U(1)_X$, promoted to local, as well as $U(1)_{PQ}$, $U(1)_{B}$ and $U(1)_{L}$, where the last two are the baryon and lepton number symmetries. For instance, let us choose the generator basis to be $(q^\prime_{n_L}, q^\prime_{Q_L}, q^\prime_{\Phi_u}, q^\prime_{\varphi}, q^\prime_{\sigma})$. In this case, the symmetries $U(1)_L$ and $U(1)_B$ are generated by $(1,0,0,0,0)$ and $(0,1/3,0,0,0)$, respectively. For $U(1)_Y$, the generator can be identified as $(0,1/6,1/2,0,0)$, which is clearly linearly independent, but not necessarily orthogonal, with respect to the previous two generators. The generators of the last two symmetries, $U(1)_X$ and $U(1)_{PQ}$, are also linearly independent among themselves and with respect to the other three. By construction, the generator of $U(1)_X$ has a non-zero fourth and a zero fifth entry: $(X_{n_L},  X_{Q_L}, X_{\Phi_u}, X_{\varphi}\neq 0, X_{\sigma}=0)$, whereas $U(1)_{PQ}$ is the only symmetry for which the last entry must be different from zero: $(PQ_{n_L},  PQ_{Q_L}, PQ_{\Phi_u}, PQ_{\varphi}, PQ_{\sigma}\neq 0)$. The exact charges that define these generators will be properly derived in the next section once the scalar spectrum is obtained, in particular, when the orthogonal Goldstone bosons are identified. This procedure allows for the unambiguous identification of the physical charges, preventing any misleading choice \cite{Quevillon:2020hmx}. Finally, the $U(1)_{afree}$ column represents a subgroup of $U(1)_{global}$, containing only anomaly-free solutions. To find $U(1)_{afree}$, we impose that all the coefficients arising from the anomalies below must vanish 
\begin{equation}\label{eq:anomalies}
    \begin{split}
    I&:\,[SU(3)_C]^2\times U(1)_{global}\,;\quad II:\, [SU(2)_L]^2\times U(1)_{global}\,;\quad III:\, [U(1)_{Y}]^2\times U(1)_{global}\,;\\
    IV&:\, U(1)_{Y}\times [U(1)_{global}]^2\,;\quad\,\,\,\,\, V:\,[Grav]^2\times U(1)_{global}\,;\quad\quad\,\, VI:\, [U(1)_{global}]^3\,.
    \end{split}
\end{equation}
For instance, the vanishing of the anomaly coefficient in $I$ is achieved for $q^\prime_\sigma\equiv l_\sigma =0$. As for the coefficient $II$, in addition to the previous constraint, its vanishing requires that $q^\prime_{n_L}\equiv l_{n_L} = -3l_{Q_L}+ l_{\Phi_u}+ l_{\varphi}$. Once these two constraints are imposed, all anomaly coefficients vanish identically, as shown in Appendix \ref{app:anomalies}. Consequently, after the imposition of two constraints, the number of independent charges goes from five, in the $U(1)_{global}$ column, to only three, in the $U(1)_{afree}$ column. Such charges can be grouped in the basis $(l_{Q_L}, l_{\Phi_u}, l_\varphi)$ and be identified as the generators of $U(1)_{B-L}$: $(1/3,0,0)$, $U(1)_{Y}$: $(1/6,1/2,0)$ and $U(1)_X$: $(X_{Q_L}, X_{\Phi_u}, X_\varphi\neq 0)$.

Although the symmetries $U(1)_{Y}$, $U(1)_X$ and $U(1)_{PQ}$ are all broken spontaneously, the $U(1)_{B-L}$ symmetry will remain intact, ensuring the Dirac nature of neutrinos, whose masses are generated via a (Dirac) seesaw mechanism.

\section{Scalar sector}\label{sec:scalar}

The scalar potential can be written as
\begin{eqnarray}\label{V}
        V &=& \mu_d^2(\Phi_d^{\dagger} \Phi_d) + \mu_u^2( \Phi_u^{\dagger} \Phi_u) + \mu_\sigma^2(\sigma^* \sigma) + \mu_\varphi^2 ( \varphi^* \varphi) + \lambda_d (\Phi_d^{\dagger} \Phi_d)^2 + \lambda_u (\Phi_u^{\dagger} \Phi_u)^2 + \lambda_\sigma (\sigma^* \sigma)^2 \nn\\
        &+& \lambda_\varphi(\varphi^* \varphi)^2 +\lambda_{du} (\Phi_d^{\dagger} \Phi_d)(\Phi_u^{\dagger} \Phi_u) + \tilde{\lambda}_{du} (\Phi_d^{\dagger} \Phi_u)(\Phi_u^{\dagger} \Phi_d) + \lambda_{\sigma\varphi} (\sigma^* \sigma)(\varphi^* \varphi) + \lambda_{d\sigma} (\Phi_d^{\dagger} \Phi_d) (\sigma^* \sigma) \nn\\
        &+& \lambda_{d\varphi}   (\Phi_d^{\dagger} \Phi_d)(\varphi^* \varphi) +\lambda_{u\sigma} (\Phi_u^{\dagger} \Phi_u)(\sigma^* \sigma) + \lambda_{u\varphi} (\Phi_u^{\dagger} \Phi_u) (\varphi^* \varphi) - \left[\lambda_4(\Phi_d^\dagger\Phi_u)(\sigma\varphi) +\mathrm{h.c.}\right]\,.
\end{eqnarray}
In the limit $\lambda_4 \to 0$, the scalar potential has four independent $U(1)$ global symmetries related to the phase redefinitions allowed for each scalar field. When the $\lambda_4$ term is introduced, one of the four possible linear combinations of these symmetries is explicitly broken so that $V$ is left invariant under only three Abelian groups. Two of them can be identified with the gauged $U(1)_Y$ and $U(1)_X$ symmetries, while the remaining one is the global $U(1)_{PQ}$ symmetry. As discussed in Sec. \ref{sec:modbuild}, the $U(1)_{PQ}$ ($U(1)_X$) charges can be obtained from the $U(1)_{global}$ ($U(1)_{afree}$) column in Table \ref{tab1} when taking $q^\prime_\sigma\equiv PQ_\sigma  \neq 0$ ($l_\varphi \equiv X_\varphi \neq 0$).

In order to derive the scalar spectrum, we decompose the scalar doublets as
\bea\label{doublets}
\Phi_{u,d} = (\phi^+_{u,d},  \,\,\,\phi^0_{u,d})^T\,\,\,\,\,\,\mbox{with}\,\,\,\,\,\,\phi^0_{u,d}=\frac{v_{u,d}+s_{u,d}}{\sqrt{2}}\exp\left( i \frac{a_{u,d}}{v_{u,d}}\right)\,,
\eea
with $\sqrt{v_u^2+v_d^2}\equiv v=246$ GeV, whilst for the singlets, we have
\bea\label{singlets}
\varphi=\frac{v_{\varphi}+s_{\varphi}}{\sqrt{2}}\exp\left(i \frac{a_{\varphi}}{v_{\varphi}}\right)\,\,\,\,\,\,\mbox{and}\,\,\,\,\,\,\sigma=\frac{v_{\sigma}+s_{\sigma}}{\sqrt{2}}\exp\left(i \frac{a_{\sigma}}{v_{\sigma}}\right). 
\eea
Once all scalars acquire vevs, the following spontaneous symmetry breaking pattern takes place
\be\label{ssbpattern}
\mathcal{G}_{SM}\otimes U(1)_X (\otimes\,U(1)_{PQ}\otimes U(1)_{B}\otimes U(1)_{L}) \to SU(3)_C \otimes U(1)_{EM}(\otimes\, U(1)_B\otimes U(1)_L)\,,
\ee
where $\mathcal{G}_{SM}$ stands for the SM gauge group and the global symmetries are shown in parentheses. Notice that, as discussed in Sec. \ref{sec:modbuild}, no scalar field is charged under the accidental $U(1)_B$ and $U(1)_L$ symmetries so that they remain fully conserved. The breaking of five generators leads to four would-be Goldstone bosons, absorbed by the gauge sector, plus a pseudo-Goldstone boson, the axion, as we derive in what follows.

Substituting Eqs. (\ref{doublets}) and (\ref{singlets}) into Eq. (\ref{V}), we extract the following tadpole equations
\begin{equation}\label{tadpole}
\begin{split}
v_u \left[2 \mu_u^2+2 \lambda_u v_u^2+v_d^2 (\lambda_{du}+\tilde{\lambda}_{du})+\lambda_{u\varphi} v_\varphi^2+\lambda_{u\sigma} v_\sigma^2\right]&=\lambda_4 v_d v_\varphi v_\sigma\,,\\
v_d \left[2 \mu_d^2+2 \lambda_d v_d^2+v_u^2 (\lambda_{du}+\tilde{\lambda}_{du})+\lambda_{d\varphi} v_\varphi^2+\lambda_{d\sigma} v_\sigma^2\right]&=\lambda_4 v_u v_\varphi v_\sigma\,,\\
v_\varphi \left(2 \mu_\varphi^2+2 \lambda_{\varphi} v_\varphi^2+\lambda_{d\varphi} v_d^2+\lambda_{u\varphi} v_u^2 +\lambda_{\sigma\varphi} v_\sigma^2\right)&=\lambda_4 v_u v_d v_\sigma\,,\\
v_\sigma \left(2 \mu_\sigma^2+2 \lambda_\sigma v_\sigma^2+\lambda_{d\sigma} v_d^2+\lambda_{u\sigma} v_u^2+\lambda_{\sigma\varphi} v_\varphi^2\right)&=\lambda_4 v_u v_d v_\varphi\,.
\end{split}
\end{equation}

To find the physical spectrum, we solve the equations above for the dimensionful parameters $\mu_u$, $\mu_d$, $\mu_\varphi$ and $\mu_\sigma$, and plug them back into the potential.

The scalar spectrum contains two charged fields, which are defined in terms of $(\phi^\pm_u\,,\phi^\pm_d)$ as
\begin{equation}\label{chsc}
    \begin{split}
        \phi^\pm = \frac{1}{v}(v_u \phi_d^\pm-v_d \phi_u^\pm) \quad &\Rightarrow \quad m_{\phi^{\pm}} = \frac{v^2}{2v_u v_d} \left(\lambda_4 v_\sigma v_\varphi-\tilde{\lambda}_{du}v_u v_d\right) \,, \\
        G^\pm = \frac{1}{v}(v_d \phi_d^\pm + v_u \phi_u^\pm) \quad &\Rightarrow \quad m_{G^\pm} = 0 \,.
    \end{split}
\end{equation}
The first field, $\phi^\pm$, is a physical charged scalar, whose mass can be around the electroweak scale, as in 2HDMs, as long as $\lambda_4$ remains very small. The smallness of such a parameter is naturally protected since in its absence the potential exhibits an enhanced symmetry. The second scalar, $G^\pm$, which remains massless, is the Goldstone boson absorbed by the gauge sector, making the SM vector boson $W^\pm$ massive.

For the neutral fields, we divide them into the CP-even and the CP-odd components. Starting with the CP-even scalars, in the basis $(s_u\,,s_d\,,s_\varphi\,,s_\sigma)$, we can write the squared mass matrix below
\be\label{CPevenMM}
M_s^2=\left(
\begin{array}{cccc}
 2 \lambda_u v_u^2+\frac{\lambda_4 v_d v_\varphi v_\sigma}{2 v_u} & \star & \star & \star \\
 v_u v_d (\lambda_{du}+\tilde{\lambda}_{du})-\frac{\lambda_4 v_\varphi v_\sigma}{2} & \frac{\lambda_4 v_u v_\varphi v_\sigma}{2 v_d}+2 \lambda_d v_d^2 & \star & \star \\
 \lambda_{u\varphi} v_u v_\varphi-\frac{\lambda_4 v_d v_\sigma}{2} & \lambda_{d\varphi} v_d v_\varphi-\frac{\lambda_4 v_u v_\sigma}{2} & \frac{\lambda_4 v_u v_d v_\sigma}{2 v_\varphi}+2 \lambda_{\varphi} v_\varphi^2 & \star \\
 \lambda_{u\sigma} v_u v_\sigma-\frac{\lambda_4 v_d v_\varphi}{2} & \lambda_{d\sigma} v_d v_\sigma-\frac{\lambda_4 v_u v_\varphi}{2} & \lambda_{\sigma\varphi} v_\varphi v_\sigma-\frac{\lambda_4 v_u v_d}{2} & \frac{\lambda_4 v_u v_d v_\varphi}{2 v_\sigma}+2 \lambda_{\sigma} v_\sigma^2 \\
\end{array}
\right).
\ee
The matrix in Eq. (\ref{CPevenMM}) contains the three energy scales present in the model, and it is expected that two scalars will be heavy  with masses proportional to the vevs $v_\varphi$ and $v_\sigma$. It is a typical feature of the axion models that the mass matrix of the CP-even scalars contains hierarchical vacuum expectation values. To have a Higgs boson consistent with the observed one, an adjustment of the parameters is required. We will not develop it further once this is not the focus of the present work.

\subsection{CP-odd sector: identifying Goldstone bosons and abelian charges}

As a result of the polar parametrisation in Eqs. (\ref{doublets}) and (\ref{singlets}), the terms in the potential involving only the CP-odd states can be succinctly written as
\be\label{CPoddV}
V(a_i) = -\frac{\lambda_4}{2} v_u v_d v_\varphi v_\sigma \cos\left(\frac{a_u}{v_u}-\frac{a_d}{v_d}+\frac{a_\varphi}{v_\varphi}+\frac{a_\sigma}{v_\sigma}\right)\,.
\ee 
Upon expanding the cosine function, we find that only one state becomes massive at this point. The massive state is proportional to the argument of the cosine function, which, when normalised, translates to 
\be\label{Astate}
A = \frac{1}{\sqrt{v_\sigma^2(v_\varphi^2 v^2 + v_u^2 v_d^2) + v_\varphi^2 v_u^2v_d^2}}\left(v_d v_\varphi v_\sigma a_u-v_u v_\varphi v_\sigma a_d+v_u v_d v_\sigma a_\varphi +v_u v_d v_\varphi a_\sigma\right),
\ee
with a squared mass given by
\be\label{mA}
m_A^2 = \frac{\lambda_4}{2}\left[\frac{v_\sigma^2(v_\varphi^2 v^2 + v_u^2 v_d^2) + v_\varphi^2 v_u^2v_d^2 }{v_u v_d v_\varphi v_\sigma}\right] \,.
\ee
The pseudoscalar field $A$ becomes massless in the limit $\lambda_4\rightarrow0$. In fact, as mentioned below Eq. (\ref{chsc}), the absence of the term $\lambda_4(\Phi_d^\dagger\Phi_u)(\sigma\varphi) +\mathrm{h.c.}$ implies the existence of an extra global symmetry in the scalar potential whose spontaneous breaking would identify the field $A$ as the associated Goldstone boson. Thus, under the assumption that $\lambda_4$ can be naturally small ($\lambda_4 \ll 1$), since its vanishing increases the symmetries of the system, $m_A$ could also be around the electroweak scale, for example. Moreover, according to the vev hierarchy, $A$ couples predominantly to the Standard Model fields once its components are mainly along the $a_u$ and $a_d$ field space directions.

As for the remaining fields, they are Goldstone bosons associated with the spontaneous breaking of three abelian symmetries: $U(1)_Y$, $U(1)_X$ and $U(1)_{PQ}$, defining a degenerate 3-d space. In order to identify the three linearly independent CP-odd scalars, it is instructive to write down the conserved current associated with each $U(1)$ symmetry in the model along the CP-odd scalars. As usual, we assume that under a given global $U(1)_c$ symmetry, a scalar field $\phi$ transforms as $\phi\rightarrow \exp(i \omega_c c_\phi) \phi$, where $c_\phi$ is $\phi$'s $U(1)_c$ charge, and $\omega_c$ is the infinitesimal continuous parameter of $U(1)_c$. Noether's theorem tells us that the presence of a $U(1)_c$ symmetry -- in our case $c=Y,X, PQ$ -- implies the conservation of the following current
\be\label{current}
J_\mu^c = \sum_\phi\frac{\partial \mathcal{L}}{\partial(\partial_\mu \phi)} \frac{\delta \phi}{\delta \omega_c} +\cdots =- i\sum_\phi c_\phi \phi^\dagger \partial_\mu \phi +\mbox{h.c.} +\cdots \,,
\ee 
where the ellipsis corresponds to the contributions from all non-scalar fields charged under the symmetry. Using the polar decomposition for the scalars, as in Eq. (\ref{doublets}) and (\ref{singlets}), we find the conserved current along the CP-odd scalars to be
\be\label{currentCPodd}
J_\mu^c|_{a_\phi} = \sum_\phi c_\phi v_\phi \partial_\mu a_\phi = f_G \partial_\mu G_c \,,
\ee 
where, in the last step, we have defined the linear combination
\be\label{Goldstone}
G_c = \frac{1}{f_G} \sum_\phi c_\phi v_\phi a_\phi, 
\ee
with $f_G^2 =\sum_\phi c^2_\phi v_\phi^2$. The field $G_c$ is precisely the massless field associated with the spontaneously broken $U(1)_c$ generator as predicted by Goldstone's theorem ($\langle 0|J^c_\mu|G_c\rangle = ip_\mu G_c$).

We are now well equipped to determine the Goldstone bosons by applying the expression in Eq. (\ref{Goldstone}) to our model. Moreover, by imposing the physical condition of orthogonality among the CP-odd states, we are able to fix the U(1) charges of the scalars in terms of the vevs, ensuring that the charges in Table \ref{tab1} are unambiguously chosen.

The first (would-be) Goldstone boson, $G_Z\equiv G_Y$, comes from the breaking of $U(1)_Y$ and is absorbed by the massive vector boson $Z$ via the Higgs mechanism. As only the $SU(2)_L$ doublets carry hypercharge, we can use Eq. (\ref{Goldstone}) to obtain, as expected,
\be\label{GZ}
G_{Z} =  \frac{1}{v} \left( v_u a_u+ v_d a_d \right) \,.
\ee 
Notice that $G_Z$ is automatically orthogonal to $A$, as it should be.
Had we not known beforehand the hypercharges of the doublets, we could also have identified Eq. (\ref{GZ}) from its orthogonality to $A$, which would in turn provide us with the relation $Y_{\Phi_u} =Y_{\Phi_d}$.

A second would-be Goldstone boson, $G_{Z^\prime}\equiv G_X$, emerges when the gauged $U(1)_X$ symmetry is spontaneously broken. $G_{Z^\prime}$ can be properly identified by noticing that it has components along the scalars charged under $U(1)_X$, {\it i.e.} $\Phi_{u,d}$ and $\varphi$ (see Table \ref{tab1}), as well as it must be orthogonal to $A$ and $G_Z$, giving
\be\label{GZprime}
G_{Z^\prime} = \frac{1}{\sqrt{v_\varphi^2 +\frac{v_u^2 v_d^2}{v^2}}}\left(-\frac{ v_d^2 v_u}{v^2}a_u+\frac{v_u^2 v_d}{v^2}a_d + v_\varphi a_\varphi\right) \,.
\ee 
By comparing Eq. (\ref{GZprime}) and (\ref{Goldstone}), we find the unambiguous $U(1)_X$ charge relations:
\be\label{Xcharges} 
\frac{X_{\Phi_u}}{X_{\varphi}} = -\frac{v_d^2}{v^2} \quad\quad\mbox{and}\quad\quad \frac{X_{\Phi_d}}{X_{\varphi}} = \frac{X_{\Phi_u}}{X_{\varphi}}+1= \frac{v_u^2}{v^2} \,.
\ee
Without loss of generality, we normalise the $U(1)_X$ charges by setting: $X_\varphi=1$.

We would like to emphasise that once the orthogonality among the Goldstone bosons is imposed, the $U(1)_X$ charges of the scalars in Table \ref{tab1} cannot be chosen freely. This feature leads to a very distinctive implication to the extended gauge sector phenomenology: no tree-level mass mixing between the SM and $U(1)_X$ gauge bosons is generated, as discussed in the next section. 

Finally, we can proceed to the last CP-odd state, the (pseudo-)Goldstone of the anomalous $U(1)_{PQ}$ symmetry, the axion $a\equiv G_{PQ}$, which can be easily obtained by requiring it to be orthogonal to $A$, $G_Z$ and $G_{Z^\prime}$: 
\be\label{axion1}
a =\frac{1}{\sqrt{ v_\sigma^2+\frac{v_u^2 v_d^2 v_\varphi^2}{v_u^2v_d^2+v_\varphi^2v^2}}}\left( -\frac{v_d^2 v_\varphi^2 v_u}{v_u^2v_d^2+v_\varphi^2v^2} a_u + \frac{v_u^2v_\varphi^2 v_d }{v_u^2v_d^2+v_\varphi^2v^2} a_d - \frac{v_u^2 v_d^2 v_\varphi}{v_u^2v_d^2+v_\varphi^2v^2} a_\varphi + v_\sigma a_\sigma\right) \,.
\ee
Once again, with the aid of Eq. (\ref{Goldstone}), we identify the $PQ$ charges of the scalars in terms of the vevs:
\be\label{PQcharges1}
\frac{PQ_{\Phi_u}}{PQ_{\sigma}} = -\frac{v_d^2v_\varphi^2}{v_u^2v_d^2+v_\varphi^2v^2} \,,\quad\quad\frac{PQ_{\Phi_d}}{PQ_{\sigma}} = \frac{v_u^2v_\varphi^2}{v_u^2v_d^2+v_\varphi^2v^2}\quad\quad\mbox{and}\quad\quad\frac{PQ_{\varphi}}{PQ_{\sigma}} = -\frac{v_u^2v_d^2}{v_u^2v_d^2+v_\varphi^2v^2} \,,
\ee 
and, as a normalisation condition, we can adopt $PQ_\sigma=1$.

Alternatively, the axion field above can be also identified by adding an explicit $U(1)_{PQ}$-breaking term of the form $\kappa \sigma^n + \mbox{h.c.}$ to the potential. Since $\sigma$ is not charged under the other Abelian symmetries, see Table \ref{tab1}, $U(1)_Y$ and $U(1)_X$ remain conserved. With the introduction of the new term, the CP-odd spectrum would contain two massive fields, one of which gets a mass proportional to $\kappa$. Then, in the limit $\kappa \to 0$ -- {\it i.e.} recovering the $U(1)_{PQ}$-invariant potential in Eq. (\ref{V}) -- the $\kappa$-dependent mass goes to zero, so that the associated state can be identified with the Goldstone boson emerging from the spontaneous breaking of $U(1)_{PQ}$. Such a field is exactly the axion given by Eq. (\ref{axion1}). We will return to the axion field in Sec. \ref{sec:axion} to study its main properties and phenomenological features.

\section{Gauge sector: unmixed $U(1)_X$ gauge boson}\label{sec:gaugespec}

The relevant terms giving rise to gauge boson masses are
\begin{equation}
    \mathcal{L} = \sum_{i=u,d} \left( D^{\mu} \Phi_i \right)^{\dagger} (D_{\mu} \Phi_i) + (D^{\mu}\varphi)^{\dagger}(D_{\mu}\varphi) \,,
\end{equation}
where the covariant derivatives are given by
\begin{equation}\label{covder}
\begin{split}
     D^{\mu} \Phi_i &= \left(\partial^{\mu} - i g_L \bm{T} \cdot \bm{W}^{\mu} - i g_Y Y_{\Phi_i} B_Y^\mu - i g_X X_{\Phi_i} B_X^\mu \right) \Phi_i \, , \quad i = u, d \,,\\
     D^{\mu}\varphi &= \left( \partial^{\mu} - i g_X X_{\varphi} B_X^\mu \right) \varphi \,.
\end{split}
\end{equation}
Notice that the kinetic term for $\sigma$ has being omitted since $\sigma$ carries no local charge, {\it i.e.} $D_\mu \sigma = \partial_\mu \sigma$.

When the scalar fields acquire a vev, the gauge bosons become massive via the Higgs mechanism. The charged gauge boson, $W_\mu^\pm = \frac{1}{\sqrt{2}} (W^1_\mu \mp i W^2_\mu)$, whose associated would-be Goldstone boson, $G^\pm$, is defined in Eq. (\ref{chsc}), gets the mass $m_{W^\pm}=g v/2$. The neutral gauge bosons, in the basis ($W_3^{\mu}$, $B_Y^{\mu}$, $B_X^{\mu}$), share the following squared mass matrix 
\be\label{NGBMM}
M^2_{NGB}=\frac{1}{4}\left(
\begin{array}{ccc}
 g_L^2 v^2 & - g_L g_Y v^2 & -2 g_L g_X v^2 \mathcal{A} \\
 - g_L g_Y v^2 & g_Y^2 v^2 & 2g_X g_Y v^2 \mathcal{A} \\
 -2 g_L g_X v^2 \mathcal{A} & 2g_X g_Y v^2 \mathcal{A} & 4 g_X^2v_\varphi^2 \mathcal{B} \\
\end{array}
\right) \,,
\ee
with the dimensionless parameters $\mathcal{A}$ and $\mathcal{B}$ given by
\begin{equation}
    \begin{split}
        \mathcal{A} &= \frac{X_\varphi}{v^2}\left( \frac{X_{\Phi_u}}{X_\varphi} v_u^2 + \frac{X_{\Phi_u}+X_\varphi}{X_\varphi}v_d^2 \right) = 0 \,, \\
    \mathcal{B} &= \frac{X_\varphi^2}{v_\varphi^2}\left( \frac{X_{\Phi_u}^2}{X_\varphi^2} v_u^2 + \frac{(X_{\Phi_u}+X_\varphi)^2}{X_\varphi^2} v_d^2 + X_\varphi^2 v_\varphi^2 \right)=\frac{1}{v^2 v_\varphi^2}\left( v_u^2 v_d^2 + v^2 v_\varphi^2 \right) \,,
    \end{split}
\end{equation}
where to obtain the right-hand sides of the equations above we have used the charges in Eq. (\ref{Xcharges}) and the normalisation condition $X_\varphi = 1$, which follow from the imposition of orthogonality among the Goldstone bosons. Due to the vanishing of $\mathcal{A}$, the mass matrix in Eq. (\ref{NGBMM}) becomes block diagonal. The upper $2\times 2$ block mixes $W_3^\mu$ and $B_Y^\mu$ and is precisely what one gets in the SM, so that its diagonalisation generates the massless photon field, $A_\mu$, and the massive $Z_\mu$. On the other hand, the $U(1)_X$ field, $Z^\prime_\mu\equiv B_{X\mu}$, remains unmixed\footnote{For simplicity, we are neglecting the kinetic mixing term which could also lead to $Z$-$Z^\prime$ mixing. Nevertheless, loop-suppressed contributions are expected to appear. See {\it e.g.} \cite{Williams:2011qb}, for a discussion on the topic.}, and its mass is
\be\label{ZprimeM}
m_{Z^\prime} = g_X \left(v_\varphi^2 + \frac{v_u^2v_d^2}{v^2}\right)^{1/2} \,.
\ee 
The interesting observation that $Z^\prime$ remains unmixed is not exclusive of our construction. In fact, we expect this to happen in other $U(1)_X$ gauge extensions once one imposes that all Goldstone bosons must be orthogonal. This, however, shall be explored elsewhere.

From now on, we assume that $Z^\prime$ is a heavy vector boson with $m_{Z^\prime}/g_X \simeq v_\varphi = 10$ TeV. With this benchmark choice, our model's predictions evade the current collider constraints on $m_{Z^\prime}$, obtained from the analysis of dilepton final states at the LHC \cite{Sirunyan:2018exx, Aad:2019fac}. Cosmological constraints related to the effective number of extra relativistic species $\Delta N_{eff} \leq 0.285$ \cite{Aghanim:2018eyx} are also relevant when selecting this benchmark. The cosmological constraint on $\Delta N_{eff}$ can be translated into a lower limit on $m_{Z^\prime}/g_X$ ($v_\varphi$) since the light right-handed neutrinos, $\nu_R$, may thermalise with the SM fields in the early universe, via $Z^\prime$-mediated interactions, and then contribute to $\Delta N_{eff}$. Whilst a detailed calculation of $\Delta N_{eff}$ is beyond the scope of the present work, we do not expect that our model's contribution to $\Delta N_{eff}$ will vary greatly with respect to that in Ref. \cite{FileviezPerez:2019cyn} whose analysed model, a gauged $U(1)_{B-L}$ construction, shares important features with ours. Thus, the results in Ref. \cite{FileviezPerez:2019cyn} have also been taken into account when selecting the benchmark $v_\varphi = 10$ TeV.

\section{Fermion sector}\label{sec:fermion}

We now turn our attention to the fermion sector, starting with the Yukawa interactions. The field content and its symmetry transformations, as shown in Table \ref{tab1}, allow us to write the following renormalisable Yukawa Lagrangian: 
\begin{equation}\label{Yuk}
    \begin{split}
            - \mathcal{L}_y &= y_{ab}^u \,\overline{Q_{aL}}\, \widetilde{\Phi_u}\, u_{bR} + y_{ab}^d\, \overline{Q_{aL}} \,\Phi_d\, d_{bR} +  y_{ab}^k \,\varphi^*\,\overline{k_{aL}}\,k_{bR} + \frac{y^\mu_{ab}\mu}{\sqrt{2}}\, \overline{k_{aL}} d_{bR} \\
    &+ y_{ab}^e\,\overline{L_{aL}}\, \Phi_u\, e_{bR} + y_{ab}^{n}\, \overline{L_{aL}}\, \widetilde{\Phi_d}\, n_{bR} + y_{ab}^{\alpha}\, \varphi\, \overline{n_{aL}} \, \nu_{bR} +  y_{ab}^{\beta} \,\sigma \,\overline{n_{aL}}\, n_{bR} + \mathrm{h.c.} \,.
    \end{split}
\end{equation}
The structure of the Yukawa Lagrangian is similar to that of the so-called flipped or type-Y 2HDM in that the $e_R$ and $u_R$ couple to the same Higgs doublet: $\Phi_u$. Once the scalar fields acquire vevs, according to Eqs. (\ref{doublets}) and (\ref{singlets}), all fermions become massive, as detailed in the next subsections. 

\subsection{Charged fermion spectrum}

We start by noticing that, similar to the flipped 2HDM, both the charged leptons and the up-type quarks get masses proportional to $\langle \Phi_u \rangle=v_u/\sqrt{2}$. The charged lepton masses can be obtained from
\be\label{clMM}
M^{e} = \frac{y^e v_u}{\sqrt{2}} \,, 
\ee 
a $3 \times 3$ matrix, which can be diagonalised by performing the bi-unitary transformation: $(U^e_L)^\dagger M^e U^e_R = \mbox{diag}(m_e, m_\mu, m_\tau)$. 
Likewise, the $3 \times 3$ up-type quark mass matrix is given by
 \be\label{upMM}
 M^{u} = \frac{y^u v_u}{\sqrt{2}} \,,
 \ee
and its diagonalisation follows from the bi-unitary transformation $(U^u_L)^\dagger M^{u} U^u_R = \mbox{diag}(m_u, m_c, m_t)$.

The remaining quarks, $d_{aL,R}$ and $k_{aL,R}$, when put together in the basis $D_{L,R}=(d, k)_{L,R}$, share the following $6 \times 6$ mass matrix
\be\label{downMM}
M^D = \frac{1}{\sqrt{2}}\begin{pmatrix}
y^d v_d & 0 \\
y^\mu \mu & y^{k} v_{\varphi}
\end{pmatrix} \,,
\ee 
where each element corresponds to a $3 \times 3$ block. The diagonalisation of $M^{D}$ follows from the bi-unitary transformation $(U^D_L)^\dagger M^{D} U^D_R = \mbox{diag}(m_d, m_s, m_b, M_1, M_2, M_3)$, with the $U^D$ matrices being $6\times 6$. At leading order, the first three masses ($m_i$) are proportional to the scale $v_d$, while the remaining masses ($M_i$) are proportional to $v_\varphi$. The diagonalisation of $M^D$ can be performed using different ansatze for the unitary matrices $U_{L,R}^D$ \cite{Schechter:1981cv, Grimus:2000vj, Hettmansperger:2011bt,Korner:1992zk}. Here, we adopt the ansatz in Refs. \cite{Grimus:2000vj, Hettmansperger:2011bt}, which allows us to approximate the unitary matrices $U_{L,R}^D$ as
\begin{equation}
\label{UFLR}
     U_{L,R}^D \equiv R_{L,R}^D  \mathcal{V}_{L,R}^D \approx \begin{pmatrix}
    \left( 1 - \frac{1}{2}B^D B^{D \dagger} \right) V^{d} & B^D V^{k} \\
    -B^{D \dagger} V^{d} &  \left( 1 - \frac{1}{2} B^{D \dagger} B^D  \right) V^{k}
    \end{pmatrix}_{L,R} \, ,
\end{equation}
where the $3 \times 3$ matrices $B^{D}_L \propto (\mu v_d)/v_\varphi^2$ and $B^{D}_R \propto \mu/v_\varphi$ -- up to the first order in $v_\varphi\gg \mu, v_d$ -- are obtained in Appendix \ref{app:block-diagon}. Assuming the benchmark $(v_d, \mu, v_\varphi) = (10^2, 10^3, 10^4) $ GeV, we have that these matrices come with the following suppression factors: $B^{D}_L \propto 10^{-3}$ and $B^{D}_R \propto 10^{-1}$.

\subsection{Neutrino spectrum: Dirac seesaw mechanism}

In this section, we show how neutrinos get naturally small masses via a type-I Dirac seesaw mechanism, which receives contributions from all the three energy scales present in the model. 

The neutral lepton masses come from the last three terms in Eq. (\ref{Yuk}). When writing $N_{L,R}=(\nu, n)_{L,R}$ as the basis, we have the following $6\times 6$ mass matrix  
\begin{eqnarray}\label{ssmm}
 M^{N} =\frac{1}{\sqrt{2}}\begin{pmatrix} 
       0 & y^n v_d \\ y^{\alpha} v_\varphi & y^{\beta} v_\sigma
      \end{pmatrix} \,,
\end{eqnarray}
with each element representing a $3\times 3$ block.

The mass terms above are strictly of a Dirac type since $U(1)_{B-L}$ is conserved. This can be easily understood by noticing that, as discussed in Sec. \ref{sec:modbuild}, the field charges under $U(1)_{B-L}$ are obtaining by taking $l_{Q_L} = 1/3$ and $l_{\Phi_u}=l_{\varphi}=0$ in the last column of Table \ref{tab1}. Therefore, because no scalar field is charged under $U(1)_{B-L}$, this symmetry is not broken spontaneously, implying that neutrinos, as well as all the other fermions, are necessarily Dirac fermions. 

\begin{figure}[h]
\begin{center}
\includegraphics[scale=1.2]{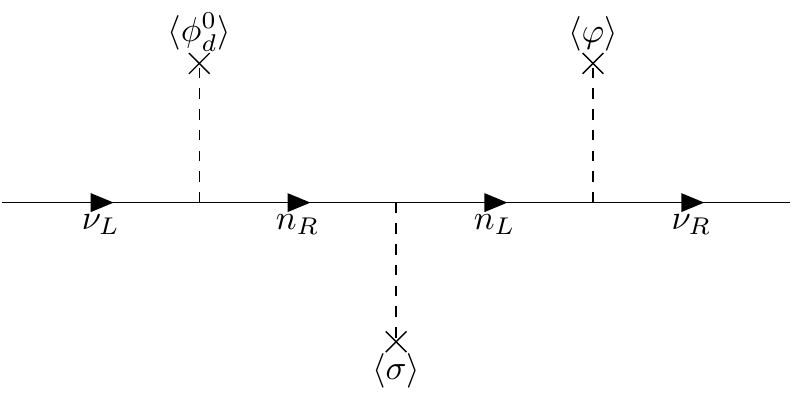}
\caption{Dirac seesaw diagram: neutrino masses suppressed by ratio of the $U(1)_X$ and the $U(1)_{PQ}$ breaking scales: $v_\varphi/v_\sigma$.}
\label{fig:DSS}
\end{center}
\end{figure}

The texture of the Dirac mass matrix $M^N$ and the assumed vev hierarchy ($v_{\sigma}\gg v_{\varphi} \gg v_{d}$) indicate that a type-I seesaw mechanism is in place.
The diagonalisation of $M^N$ is achieved by a bi-unitary transformation  $(U_L^N)^\dagger M^N(U_R^N)=\mbox{diag}(m_{\nu^\prime_1}, m_{\nu^\prime_2}, m_{\nu^\prime_3}, m_{n^\prime_1},m_{n^\prime_2},m_{n^\prime_3})$, which can be divided into two steps by writing $U_{L,R}^N = R_{L,R}^N \mathcal{V}_{L,R}^N$ \cite{Grimus:2000vj,Hettmansperger:2011bt}, given in Appendix \ref{app:block-diagon}, similar to what has been done with the $U_{L,R}^D$ matrices in Eq. (\ref{UFLR}). When the first transformation (with $R_{L,R}^N$) is performed, we obtain the seesaw-suppressed mass matrix for the active neutrinos
\begin{equation} \label{nuMM}
m_\nu \simeq \frac{ y^n (y^\beta)^{-1} (y^{\alpha})^T }{\sqrt{2}}   \frac{v_d v_\varphi}{v_\sigma} = \frac{Y^\nu_{eff}}{\sqrt{2}} \frac{v_d v_\varphi}{v_\sigma}\,,
\end{equation}
whose diagram is shown in Fig. \ref{fig:DSS}. Small neutrino masses $\lesssim 0.1$ eV can be naturally obtained when the $U(1)_X$-breaking scale is much smaller than the PQ scale: $v_\varphi/ v_\sigma\ll 1$. For instance, taking $v_d = 10^2$ GeV, $v_\varphi = 10^4$ GeV and  $v_\sigma = 10^{12}$ GeV, sub-eV neutrino masses are obtained for $ Y^\nu_{eff} \lesssim 10^{-4}$. 

For large $v_\sigma$, as in our benchmark above, the associated invisible axion can also account for the observed dark matter relic density. Therefore, the origin of small neutrino masses, a solution for the strong CP problem and the nature of dark matter go hand in hand in our construction. To illustrate the viability of our model, in Sec. \ref{sec:axion} (see Fig. \ref{fig:anuplot}), we identify a parameter space region within which the above-mentioned issues can be solved and that can be probed by forthcoming experimental searches. Furthermore, small neutrino masses, being proportional to $v_\varphi/v_\sigma$, also rely on the existence of a moderate $U(1)_X$ scale ($v_\varphi = 10$ TeV). Therefore, TeV-scale $U(1)_X$ signatures, mediated by {\it e.g.} the extra gauge boson, $Z^\prime$, may also be within the reach of current or near-future experiments, such as the high-luminosity LHC.

\subsection{Fermion couplings to vector bosons}\label{subsec:fgint}

In the previous sections, we have shown that our construction extends the SM field content not only by adding an extra gauge boson but also extra neutral leptons ($n_{L,R}, \nu_R$) and down-type quarks ($k_{L,R}$) which mix with their SM siblings. In what follows, we show that, due to these extra ingredients, fermions and vector bosons couple in a non-standard way.

The kinetic Lagrangian for fermions, giving rise to the fermion-vector boson interactions, can be, as usual, represented by
\be\label{kinfer}
\mathcal{L}_{DF} = \sum_{F,j} \overline{F_j} i \gamma_\mu D^\mu F_j\,,
\ee 
where $F$ spans through all the fermion fields in Table \ref{tab1} and $j$ through their generations, and $D^\mu$ is the covariant derivative, as in Eq. (\ref{covder}). 

Let us start by describing the fermion couplings to the massive neutral vector bosons $Z$ and $Z^\prime$, which remain unmixed at tree level, as discussed in Sec. \ref{sec:gaugespec}. Upon expanding the covariant derivative terms and transforming the fields to their mass bases, we can write the fermion couplings to the massive neutral vector bosons as
\begin{equation}\label{fermionZs}
  \mathcal{L}_{N.C.}=  \sum_{\tilde{Z}, F^\prime}\sum_{j,l} \tilde{Z^\mu}\, \overline{F^\prime_j}\gamma_\mu \left[g_{\tilde{Z} F^\prime}^V - g_{\tilde{Z} F^\prime}^A\gamma^5\right]_{jl}F^\prime_l\,,
\end{equation}
where $\tilde{Z}= Z, Z^\prime$, and $F^\prime = e^\prime, N^\prime, u^\prime$ and $D^\prime$ are the mass states defined in the previous sections. Since the extra neutral leptons and down-type quarks mix with the SM fields, flavour-changing neutral currents (FCNCs), mediated by both $Z$ and $Z^\prime$, appear and are governed by the factors 
\be\label{XFLR}
\mathcal{X}^{F_{L,R}} = \left(U^{F}_{L,R}\right)^\dagger \mbox{diag}\left(0_{3}, \mathcal{I}_{3}\right) U^{F}_{L,R} \,,
\ee 
where $U^F_{L,R}$ are the $6\times 6$ unitary matrices that diagonalise the generalised down-type quark and neutral lepton mass matrices in Eq. (\ref{downMM}) and (\ref{nuMM}), respectively, and $\mathcal{I}_3$ ($0_{3}$) is the $3\times 3$ identity (zero) matrix. The full vector and axial couplings are presented in Table \ref{tab2}, in which we used the $U(1)_X$ charge definitions in Eq. (\ref{Xcharges}), with the normalisation $X_\varphi=1$, as well as $X_{Q_L}=1/3$.

\begin{table}[h]
\begin{tabular}{|c|c|c|c|c|}
\hline
          & $e^\prime$ & $N^\prime = (\nu^\prime\,,\, n^\prime)$ & $u^\prime$ & $D^\prime =(d^\prime\,,\, k^\prime)$              \\\hline\hline
\,\,$\tilde{g}^V_{Z}$\,\,  & $(4\sin^2\theta_W-1)\mathcal{I}_{3}$   & $\mathcal{I}_{6}-\mathcal{X}^{N_L}$ & $(1-\frac{8}{3}\sin^2 \theta_W)\mathcal{I}_{3}$ & $(\frac{4}{3}\sin^2 \theta_W - 1)\mathcal{I}_{6} + \mathcal{X}^{D_L}$     \\\hline
$\tilde{g}^A_{Z}$  & $-\mathcal{I}_{3}$  & $\mathcal{I}_{6}-\mathcal{X}^{N_L}$    & $\mathcal{I}_{3}$ & $-\mathcal{I}_{6} + \mathcal{X}^{D_L}$  \\\hline
$\tilde{g}^V_{Z^\prime}$  & $\left(\frac{v_d^2}{v^2}-2\right)\mathcal{I}_{3}$  & $ -\left(2 +\frac{v_d^2}{v^2}\right)\mathcal{I}_{6} +  \frac{v_u^2}{v^2} \mathcal{X}^{N_L} + \mathcal{X}^{N_R} $  & $ \left(\frac{2}{3} - \frac{v_d^2}{v^2}\right)\mathcal{I}_{3} $ & $ \left(\frac{2}{3}   - \frac{v_u^2}{v^2}\right)\mathcal{I}_{6} - \frac{v_u^2}{v^2} \mathcal{X}^{D_L}+ \mathcal{X}^{D_R}$ \\\hline
$\tilde{g}^A_{Z^\prime}$  & $-\frac{v_d^2}{v^2}\mathcal{I}_{3}$ & $
\frac{v_d^2}{v^2}\mathcal{I}_{6} + \frac{v_u^2}{v^2}\mathcal{X}^{N_L} -  \mathcal{X}^{N_R} $ & $\frac{v_d^2}{v^2}\mathcal{I}_{3}$ & $ \frac{v_u^2}{v^2}\mathcal{I}_{6 } - \frac{v_u^2}{v^2} \mathcal{X}^{D_L} -\mathcal{X}^{D_R}$   \\\hline
\end{tabular}
\caption{Vector and axial fermion couplings to $Z$ and $Z^\prime$, where $g^{V,A}_{Z}=[g_L/(4\cos\theta_W)]\times \tilde{g}^{V,A}_{Z}$ and $g^{V,A}_{Z^\prime}=(g_X/2)\times \tilde{g}^{V,A}_{Z^\prime}$. The primed fermion fields represent the mass states, and $\mathcal{X}^{F_{L,R}}$, defined in Eq. (\ref{XFLR}), lead to flavour violation.}
\label{tab2}
\end{table}

For instance, from the coefficients in Table \ref{tab2} and Eqs. (\ref{UFLR}) and (\ref{XFLR}), we find that the most relevant source of FCNC is given by the following Lagrangian involving the (known) down-type quarks and the Z boson 
\be 
\mathcal{L}_{FCNC}^Z = \frac{g_L}{2 \cos\theta_W}Z^\mu\, \overline{d^\prime_{iL}}\gamma_\mu \left(V^{d\dagger}_L B^{D}_L B^{D\dagger}_L V^{d}_L\right)_{ij} d^\prime_{jL}  + \cdots \,, \quad \mbox{with} \quad i,j = d, s, b \,.
\ee
Nevertheless, these interactions are very suppressed in our model by the reason that, as derived in Appendix \ref{app:block-diagon}, $B^{D}_L \propto (\mu v_d)/v_\varphi^2 \simeq 10^{-3}$, where we have assumed $(v_d, \mu, v_\varphi) = (10^2, 10^3, 10^4)$ GeV. Therefore, the flavour-violating contributions come with an estimated suppression factor of at least $10^{-6}$, which is well below the current limit ($< 10^{-3}$) obtained from processes such as $B\to s\gamma$, $B_s\to \mu^+\mu^-$ and $B_s - \overline{B_s}$ mixing  \cite{AguilarSaavedra:2002kr,Vatsyayan:2020jan}. For the neutral leptons, the flavour-violating contributions, suppressed by $v_\sigma= 10^{12}$ GeV, are much smaller and can be also safely neglected.

When it comes to the charged vector bosons, $W^\pm$, we also expect corrections to the SM contributions due to the presence of the extra fermions in our model. The relevant interactions are
\begin{equation}
    \begin{split}
        \mathcal{L}_{c.c.} = \frac{g_L}{\sqrt{2}} &\left[ \overline{u^\prime_{jL}}\, \gamma^\mu \left(U_{mix}^q\right)_{jl}\, d^\prime_{lL} +  \overline{\nu^\prime_{jL}} \,\gamma^\mu \left(U_{mix}^{l\dagger}\right)_{jl} \,e^\prime_{lL} \right.\\
        &+\left.\overline{u^\prime_{jL}}\, \gamma^\mu \left(U^{u\dagger}_L F^D_L V^k_L\right)_{jl}\, k^\prime_{lL}  + \overline{n^\prime_{jL}}\, \gamma^\mu \left(V^{n\dagger}_L F^{N\dagger}_L U^e_L\right)_{jl} \,e^\prime_{lL}\right] W^+_\mu + \mbox{h.c.}\,,
    \end{split}
\end{equation}
where
\begin{equation}\label{ckmpmns}
    \begin{split}
        U_{mix}^q &= U^{u\dagger}_L \left(1-\frac{1}{2}B^D_L B^{D\dagger}_L \right)V^d_L\,,\\
        U_{mix}^{l} &= U^{e\dagger}_L\left(1-\frac{1}{2}B^N_L B^{N\dagger}_L \right)V^{\nu}_L\,,
    \end{split}
\end{equation}
are our model quark and lepton mixing matrices, oftentimes called Cabibbo-Kobayashi-Maskawa (CKM) and Pontecorvo-Maki-Nakagawa-Sakata (PMNS) matrices. Contrary to the SM case, these matrices are not unitary due to the new contributions coming from the mixing with the extra fermions encoded in $B^D_L$ and $B^N_L$. The deviations from unitarity are, however, very suppressed since they are roughly of the order $(B^D_L)^2 \sim (\mu v_d/v_\varphi^2)^2 = 10^{-6}$ and $(B^{N}_L)^2 \sim (v_d/v_\sigma)^2 = 10^{-20} $. 

Finally, for the sake of completeness, we provide the fermion couplings to photons, which, as expected, are given by 
\begin{equation}
    \mathcal{L}_{A_\mu} =  e A_\mu \sum_{j} Q_{F^\prime_j} \overline{F^\prime_j}\gamma^\mu F^\prime_j \,,
\end{equation}
where $F^\prime$ varies through all the fermion mass bases and $eQ_{F^\prime}$ represents the electric charge of $F^\prime$.

\section{Axion physics}\label{sec:axion}

In this section, we bring our attention back to the axion field and discuss some of its relevant properties. To do so, it is convenient to rewrite the axion field, according to Eq. (\ref{Goldstone}), as \cite{Srednicki:1985xd}
\begin{equation}\label{a2}
 a = \frac{1}{f_{PQ}}\left[  v_u PQ_{\Phi_u}\, a_u +  v_d PQ_{\Phi_d}\, a_d +  v_\varphi PQ_{\varphi}\, a_\varphi +  v_\sigma PQ_{\sigma}\,a_\sigma \right]\,,
\end{equation}
where the $PQ$ charges, defined in Eq. (\ref{PQcharges1}) with $PQ_\sigma=1$, as well as $f_{PQ}$, reparametrised in terms of two angles
\begin{eqnarray}\label{PQsc}
PQ_{\Phi_u} =  - (\cos{\beta}\cos{\theta})^2 \,,\quad\quad
PQ_{\Phi_d} =  (\sin{\beta}\cos{\theta})^2,\quad\quad
PQ_{\varphi} = - (\sin{\theta})^2 \,,
\end{eqnarray}
and 
\begin{eqnarray}\label{fPQ}
f_{PQ} = \sqrt{ v_\sigma^2 + v_\varphi^2\sin^4\theta + \cos^4 \theta\left(v_d^2 \sin^4\beta + v_u^2\cos^4\beta\right) }\,,
\end{eqnarray}  
where $\beta$ and $\theta$ are defined as
\begin{eqnarray}\label{thetas}
\tan{\beta} = \frac{v_u}{v_d}\quad\quad\mbox{and} \quad\quad \tan{\theta} = \frac{v_uv_d}{ v_\varphi v} \,.
\end{eqnarray}
Notice that the first angle follows the conventional definition of $\beta$ in 2HDM scenarios \cite{Branco:2011iw}, whilst $\theta$ is expected to be small due to the assumed vev hierarchy $v_\sigma \gg v_\varphi \gg v$. Thus, the axion field is predominantly projected along the CP-odd component of the singlet $\sigma$: $a\simeq a_\sigma$, and $f_{PQ}\simeq v_\sigma$. It is also worth pointing out that the charges in Eq. (\ref{PQsc}) satisfy the relation
\begin{eqnarray}\label{PQchrel}
PQ_{\Phi_u}-PQ_{\Phi_d}+PQ_{\Phi_\varphi}= -PQ_\sigma=-1 \,,
\end{eqnarray}
which arises from the only non-hermitian term in the scalar potential, {\it i.e.} $\lambda_4(\Phi_d^\dagger \Phi_u)(\sigma \varphi)$.

In order to solve the strong CP problem through the Peccei-Quinn mechanism, the $U(1)_{PQ}$ symmetry must yield a nonzero $[SU(3)_C]^2\times U(1)_{PQ}$ anomaly coefficient $C_{ag}$. This leads to the effective interaction for the axion field with the gluon field strength $G_{\mu\nu}^b$:  
\begin{eqnarray}
{\cal L}_{agg}=-\frac{\alpha_s}{8\pi}\frac{C_{ag}}{f_{PQ}}a\,G_{\mu\nu}^b\tilde{G}^{b,\mu\nu} \,,
\end{eqnarray}
in which $\alpha_s=g_s/(4\pi)$, where $g_s$ is the strong interaction coupling constant, and $\tilde{G}^{b,\mu\nu}\equiv\epsilon^{\mu\nu\sigma\rho}G_{\sigma\rho}^b/2$ is the dual field strength. As we shall see in Eq. (\ref{EandN}), the model has indeed a nonvanishing anomaly coefficient $C_{ag}=3PQ_\sigma=3$. From this, we define the axion decay constant $f_a=f_{PQ}/N_{DW}$ as well as the domain wall number $N_{DW}=C_{ag}=3$ for the model. The same domain wall number occurs in the DFSZ-type model with the non-Hermitian term $\Phi_u^\dagger \Phi_d \sigma$ in the scalar potential, but contrasts with the DFSZ-type model having instead the non-Hermitian term $\Phi_u^\dagger \Phi_d \sigma^2$, which leads to  $N_{DW}=6$ (see \cite{DiLuzio:2020wdo}, for example).

The axion mass arises from nonperturbative QCD effects~\cite{Weinberg:1977ma}. The leading order axion mass is given by 
\begin{equation}
    m_a=\frac{\sqrt{m_u m_d}}{m_u + m_d}\frac{m_\pi f_\pi}{f_a}
    \approx 5.7\left(\frac{10^{12}\,\mbox{GeV}}{f_a}\right) \mbox{$\mu$eV} \,,
\end{equation}
in which $m_u$ ($m_d$) is the up-quark (down-quark) mass, $m_\pi$ the pion mass,  $f_\pi\simeq 93$ MeV and $f_a$ the pion and the axion decay constants \cite{Weinberg:1977ma}. Corrections to this formula, including higher orders in chiral perturbation theory and through lattice simulations, were obtained in~\cite{diCortona:2015ldu,Borsanyi:2016ksw,Gorghetto:2018ocs}. 
Taking into account the benchmark $v_\sigma=10^{12}$ GeV for the $U(1)_{PQ}$ symmetry scale, defined in Eq. (\ref{fPQ}), we have that $f_a\simeq v_\sigma/N_{DW}$, which leads to the axion mass $m_a\simeq 17\,\mbox{$\mu$eV}$. 

Despite being very light, axions can account for a part of or even the total cold dark matter in the Universe, with their production realised through the vacuum re-alignment mechanism~\cite{Abbott:1982af,Preskill:1982cy,Dine:1982ah}. If the $U(1)_{PQ}$ symmetry breaking happened before or during the inflationary period of the Universe and was not restored afterwards, it is estimated that the axion field gives rise to a contribution to the dark matter relic density given by~\cite{Zyla:2020zbs,Sikivie:2020zpn,DiLuzio:2020wdo}
\begin{equation}
    \Omega_a h^{2}\approx 0.12\, F\,\theta^2_i\left(\frac{f_a}{9\times10^{11}\,\mbox{GeV}}\right)^{1.165} \,, 
\end{equation}
where $\theta_i$ is the initial misalignment angle which assumes values in the interval [$-\pi,\,\pi$], and the factor $F$ accounts for anharmonicities that may occur in the axion potential. For example, with the benchmark $f_a\simeq 3.3\times10^{11}$ GeV, the axion could comprise the totality of the observed cold dark matter, i.e. $\Omega_{CDM} h^{2}=0.12$ \cite{Aghanim:2018eyx}, if  $F\,\theta^2_i\approx3.2$. This is consistent with cosmological observations if the Hubble expansion rate during inflation satisfies the constraint $H_{inf}\lesssim10^7$ GeV \cite{Akrami:2018odb}, which follows from the non-observation of isocurvature fluctuations in the cosmic microwave background arising from quantum fluctuations of the axion field.

\subsection{Axion coupling to photons}

The interaction between an axion and two photons can be expressed as 
\be
{\cal L}_{a\gamma\gamma}=-\frac{g_{a\gamma}}{4}a\,F_{\mu\nu}\tilde{F}^{\mu\nu} \,,
\label{laxga}
\ee
where the coupling $g_{a\gamma}$ is
\be
g_{a\gamma}\approx\frac{\alpha}{2\pi f_a}\left(\frac{C_{a\gamma}}{C_{ag}}-1.95\right) \,,
\label{gag}
\ee
with $\alpha$ being the fine-structure constant. The factor $-1.95$ is model independent and comes from the ratio between the up- and down-quark masses, showing up in the calculation due to the mixing between axions and pions~\cite{Dias:2014osa}. Moreover, we have the model dependent contribution $C_{a\gamma}/C_{ag}$ -- oftentimes defined as $E/N$ in the axion literature -- where $C_{a\gamma}$ and $C_{ag}$ are, respectively, the $[U(1)_Q]^2\times U(1)_{PQ}$ and $[SU(3)_C]^2\times U(1)_{PQ}$ anomaly coefficients, defined by \begin{equation}\label{EandN}
    \begin{split}
        C_{a\gamma} &\equiv 2 \sum_{f=fermions}(PQ_{f_L}-PQ_{f_R})(Q_{f})^2 = 2PQ_\sigma \,,\\
  C_{ag} &\equiv \sum_{q=quarks}(PQ_{q_L}-PQ_{q_R}) = 3PQ_\sigma \,.
    \end{split}
\end{equation}
This result, $C_{a\gamma}/C_{ag}=2/3$, is precisely what one gets in the type-II DFSZ model. Therefore, although our model has extra quarks which contribute to both $C_{a\gamma}$ and $C_{ag}$, the ratio $C_{a\gamma}/C_{ag}$ remains the same as in the DFSZ case. 
\begin{figure}[H]
\begin{center}
\includegraphics[scale=0.44]{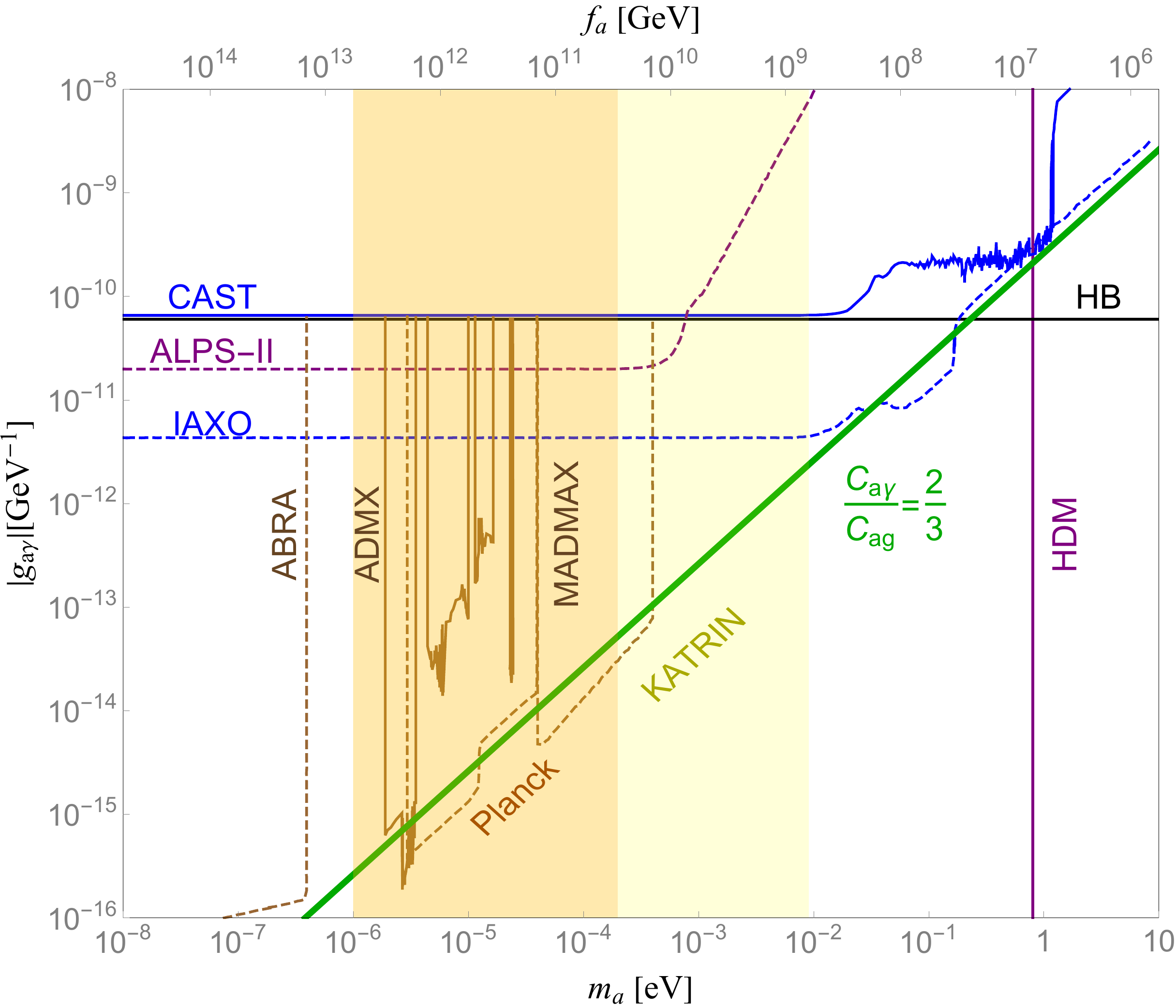}
\caption{The green line shows our model prediction for $|g_{a\gamma}|$ vs $m_a$. Constraints from experiments, cosmology and astrophysics are also displayed: the solid lines delimit exclusion regions (CAST \cite{Zioutas:2004hi,Andriamonje:2007ew,Anastassopoulos:2017ftl}, ADMX \cite{Duffy:2006aa,Asztalos:2009yp,Asztalos:2011bm,Stern:2016bbw,Braine:2019fqb}, HB bound \cite{Ayala:2014pea, Straniero:2015nvc}, HDM \cite{Hannestad:2010yi,Archidiacono:2013cha,DiValentino:2015wba}), whereas the dashed lines indicate projected experimental sensitivities (ABRACADABRA \cite{Kahn:2016aff,Ouellet:2018beu}, MADMAX \cite{TheMADMAXWorkingGroup:2016hpc}, IAXO \cite{Armengaud:2019uso}, ALPS II \cite{Hartman:2020bgj, Bahre:2013ywa, Bastidon:2015efa}). The shaded orange and orange+yellow bands identify the preferred regions for neutrino masses once the limits from the Planck Collaboration \cite{Vagnozzi:2017ovm, Aghanim:2018eyx} and KATRIN experiment \cite{Aker:2019uuj} are considered, respectively, with effective Yukawas in the range $2.9 \times 10^{-6}\lesssim Y^\nu_{eff} \lesssim 1.2\times 10^{-3}$  - see text for details. }
\label{fig:anuplot}
\end{center}
\end{figure}

In Fig. \ref{fig:anuplot}, we plot our model prediction for the axion-photon coupling as a function of the axion mass (solid green line) and also show the current constraints coming from experimental searches, cosmology and astrophysics (other solid lines) -- CAST \cite{Zioutas:2004hi,Andriamonje:2007ew,Anastassopoulos:2017ftl}, ADMX \cite{Duffy:2006aa,Asztalos:2009yp,Asztalos:2011bm,Stern:2016bbw,Braine:2019fqb}, HB bound \cite{Ayala:2014pea, Straniero:2015nvc}, HDM \cite{Hannestad:2010yi,Archidiacono:2013cha,DiValentino:2015wba} -- as well as predicted experimental sensitivities (dashed lines) -- ABRACADABRA \cite{Kahn:2016aff,Ouellet:2018beu}, MADMAX \cite{TheMADMAXWorkingGroup:2016hpc}, IAXO \cite{Armengaud:2019uso}, ALPS II \cite{Hartman:2020bgj, Bahre:2013ywa, Bastidon:2015efa}. In addition, we highlight the preferred range in the $|g_{a\gamma}|-m_a$ plane for natural neutrino masses once the limits from the Planck Collaboration \cite{Vagnozzi:2017ovm, Aghanim:2018eyx} (orange band), and KATRIN experiment \cite{Aker:2019uuj} (yellow$+$orange bands) are imposed; the relevant best fit values \cite{deSalas:2020pgw} for the normal ordering scenario are assumed, and we also take $v_d = 10^2$ GeV, $v_\varphi = 10^4$ GeV. The orange band starts at $m_a = 10^{-6}$ eV ($f_a = 5.7\times 10^{12}$ GeV) -- so that the axion field may be capable of explaining the observed DM relic density -- reproducing the lightest possible mass for the heaviest neutrino, {\it i.e.} $m_{\nu_3} = \Delta m_{31}$, for an effective Yukawa $Y^\nu_{eff} = 1.2 \times 10^{-3} $. As a criterion to select natural neutrino masses, we only consider those masses whose effective Yukawa couplings are at least of the size of the SM electron Yukawa, {\it i.e.} $(Y^\nu_{eff})_{min}\equiv Y^e_{SM} = \sqrt{2}m_e/v  \simeq 2.9 \times 10^{-6}$. The shaded region spans to the right of the plot up until the effective Yukawa reaches its minimum value, $(Y^\nu_{eff})_{min}$, and the corresponding neutrino mass reproduces the largest possible value for $m_{\nu_3}$ according to constraints coming from the Planck Collaboration (orange) -- $\sum_j m_{\nu_j} < 0.12$ eV -- and KATRIN (orange+yellow) -- $m_{ee}< 1.1$ eV. Finally, we would like to point out that, within these shaded regions, the model's prediction, {\it i.e.} the solid green line, also falls within the projected sensitivities of axion dark matter searches (MADMAX and ADMX) - dashed lines in brown.

\subsection{Axion couplings to fermions}

The charged leptons and the up-type quarks are charged universally under $U(1)_{PQ}$, and, as a result, their couplings to axions are necessarily flavour-conserving and axial and can be written as
\begin{equation}\label{af1}
 \mathcal{L}_{aF^\prime}= -i \, a\, \sum_{F^\prime,j} \overline{F^\prime_j}(g_{aF^\prime})_{j} \gamma^5 F^\prime_j \,,
\end{equation}
where $F^\prime = e^\prime, u^\prime$ and $F^\prime = F^\prime_L + F^\prime_R$ are the mass states which are associated with the flavour states via the unitary transformations $F_{L,R} = U_{L,R}^F F^\prime_{L,R}$ with indices omitted. The coefficients are given by
\begin{equation}\label{gaegau}
    \begin{split}
        (g_{ae^\prime})_j &=-\frac{\cos^2{\beta}\,\cos^2{\theta}}{3 f_a}m^{e^\prime}_j \,, \quad \mbox{with} \quad m^{e^\prime}_j= m_e, m_\mu, m_\tau \,,  \\
        (g_{au^\prime})_j &=\frac{\cos^2{\beta}\,\cos^2{\theta}}{3 f_a}m^{u^\prime}_j   \quad \mbox{and} \quad m^{u^\prime}_j= m_u, m_c, m_t \, .
    \end{split}
\end{equation}

The axion couplings to the neutral leptons and the down-type quarks bear similarities. In addition to the SM fermions, both sectors contain new fields which transform differently under the $U(1)_{PQ}$ symmetry, inducing axion-mediated FCNCs \cite{Ema:2016ops, Bjorkeroth:2018dzu}. Due to flavour-violating interactions, scalar and axial couplings appear, which can be generically written as 
\begin{equation}\label{axionFCNC}
  \mathcal{L}_{aF^\prime}=   i a\sum_{F^\prime,m,n} \overline{F^\prime_m}\left[g_{aF^\prime}^S- g_{aF^\prime}^A\gamma^5\right]_{mn}F^\prime_n \,,
\end{equation}
with $F^\prime = N^\prime, D^\prime$, and the mass basis, $F^\prime=F^\prime_L+F^\prime_R$, is related to the flavour basis via $F= U^F_{L,R}F^\prime$. 
The scalar and axial coefficients for the neutral leptons are given by
\begin{equation}
    \begin{split}
        (g_{aN^\prime}^S)_{mn} &= \frac{m^{N^\prime}_m-m^{N^\prime}_n}{6 f_a} \left[ (1+\sin^2\beta\cos^2\theta) \mathcal{X}^{N_L}_{mn} - (1+ \sin^2\theta) \mathcal{X}^{N_R}_{mn} \right] \,,\\
        (g_{aN^\prime}^A)_{mn} &= \frac{m^{N^\prime}_m+m^{N^\prime}_n}{6 f_a} \left[(\cos^2\beta\cos^2\theta-2) \delta_{mn} + (1+ \sin^2\theta) \mathcal{X}^{N_R}_{mn} \right.\\
        &\left. \quad \quad \quad \quad \quad \quad \, +  (1+\sin^2\beta\cos^2\theta) \mathcal{X}^{N_L}_{mn}  \right] \,,
    \end{split}
\end{equation}
with $\mathcal{X}^{N_{L,R}}$ defined according to Eq. (\ref{XFLR}), and the $m_n^{N^\prime}$ are the eigenvalues of the neutral lepton mass matrix in Eq. (\ref{ssmm}). As for the down-type quarks, we have
\begin{equation}
    \begin{split}
        (g_{aD^\prime}^S)_{mn} &= \frac{m^{D^\prime}_m-m^{D^\prime}_n}{6 f_a}\left[ -\sin^2\beta\cos^2\theta \mathcal{X}^{D_L}_{mn}-\sin^2\theta \mathcal{X}^{D_R}_{mn}\right] \,,\\ (g_{aD^\prime}^A)_{mn} &= \frac{m^{D^\prime}_m+m^{D^\prime}_n}{6 f_a}\left[\sin^2\beta\cos^2\theta\delta_{mn} -\sin^2\beta\cos^2\theta \mathcal{X}^{D_L}_{mn}+\sin^2\theta \mathcal{X}^{D_R}_{mn}\right] \,,
    \end{split}
\end{equation}
where $m^{D^\prime}_m$ are the masses of the down-type quarks, {\it i.e.} the eigenvalues of the mass matrix in Eq. (\ref{downMM}); $\mathcal{X}^{D_{L,R}}$ leads to FCNCs and follows from the definition in Eq. (\ref{XFLR}).

\subsection{Constraining $f_a$ with a flavour-violating process}

It is possible to set constraints on the range of the axion decay constant by confronting our model's predictions with experimental bounds on flavour-violating processes involving an axion, such as the decays of heavy mesons. The most stringent constraint comes from the process $K^{+} \to \pi^{+}a$ and is given in terms of its branching ratio: $\text{Br}(K^{+} \to \pi^{+}a) \lesssim 7.3 \times 10^{-11}$ \cite{Adler:2008zza}.

At tree level, this branching ratio is evaluated as \cite{Ema:2016ops,Bjorkeroth:2018dzu}
\begin{eqnarray}
    \label{eq:PhenomBR}
    \text{Br}(K^{+} \to \pi^{+}a) = \frac{m_K^3}{16 \pi \Gamma_{tot}} \left(1 - \frac{m_\pi^2}{m_K^2}  \right)^3 \left| \frac{(g_{aD'}^S)_{12}}{m_s-m_d} \right|^2 \, ,
\end{eqnarray}
where $\Gamma_{tot} \simeq 5.3 \times 10^{-17}$ GeV is the total decay width of $K^+$, $m_K = 493.677$ MeV and $m_\pi = 139.57$ MeV are the kaon and pion masses \cite{Zyla:2020zbs}, respectively, and the model-dependent contribution can be written as
\begin{equation}
    \label{eq:gad}
    \left|\frac{(g_{aD'}^S)_{12}}{m_s-m_d}\right| = \frac{|\sin^2 \beta \cos^2 \theta \mathcal{X}^{D_L}_{12} + \sin^2 \theta \mathcal{X}^{D_R}_{12}|}{6 f_a} \,.
\end{equation}
The relevant flavour-violating terms can be obtained from Eq. (\ref{XFLR}) and the Appendix \ref{app:block-diagon}, leading to $\mathcal{X}^{D}_{12} = (V^{d\dagger} B^{D} B^{D\dagger}V^{d})_{12}$. As derived in Appendix \ref{app:block-diagon}, the matrices $B^D$ represent the mixing between standard and new quarks and, as such, are proportional to suppression factors given in Table \ref{tab3}. By singling out the suppression factors, we can rewrite the $\mathcal{X}_{12}^D$ terms as $\mathcal{X}^{D_L}_{12}= (\frac{\mu v_d}{v_\varphi^2})^2 Y^{D_L}_{12}$ and $\mathcal{X}^{D_R}_{12}= (\frac{\mu}{v_\varphi})^2 Y^{D_R}_{12}$, where $Y^{D}_{12}$ represent all the matrix element products. Finally, we can substitute these expressions back into Eq. (\ref{eq:PhenomBR}) and when comparing it with the experimental limit, $\text{Br}(K^{+} \to \pi^{+}a) \lesssim 7.3 \times 10^{-11}$, we find that $f_a$ must satisfy
\begin{equation}
    \label{eq:fa}
    f_a \gtrsim 1.15 \times 10^{11}  \,\left( \frac{\mu^2}{v_\varphi^2} \sin^2 \theta\right) \left|Y^{D_L}_{12} + Y^{D_R}_{12}\right| \text{GeV}\,.
\end{equation}
Considering the benchmark assumed in previous sections for the scales in the model, {\it i.e.} $(v_d, \mu, v_\varphi) = (10^2,10^3,10^4)$ GeV, we find that $(\mu^2/v_\varphi^2)\sin^2 \theta \simeq 8.3\times 10^{-7}$, which leads to the weak constraint: $f_a\gtrsim 10^{5} \left|Y^{D_L}_{12} + Y^{D_R}_{12}\right| \text{GeV}$. Thus, we observe that, since the FCNC contributions arise from the mixing with non-SM heavy quarks, the constraints on $f_a$ coming from flavour-violating processes are weakened. As a result, in this construction, supernovae limits on the axion decay constant \cite{DiLuzio:2020wdo} turn out to be more stringent.

\section{Conclusions}\label{sec:conc}

We have proposed a gauged two-Higgs-doublet model featuring an axion. Dangerous tree-level FCNCs, common in 2HDMs, are forbidden by the extra $U(1)_X$ gauge symmetry. Our construction suggests that solutions for the important issues of the nature of dark matter, the origin of neutrino masses and the strong CP problem may arise from the axion-neutrino interplay.

The extended scalar sector counts with two singlets, $\sigma$ and $\varphi$, in addition to the Higgs doublets, $\Phi_d$ and $\Phi_u$. As presented in Sec. \ref{sec:scalar}, when all scalars acquire vevs, satisfying the hierarchy $v_\sigma \gg v_\varphi \gg v=(v_d^2 + v_u^2)^{1/2}$, spontaneous symmetry breaking takes place giving rise to five Goldstone bosons - four of which are, in fact, would-be Goldstone bosons absorbed by the gauge sector, while the last one is identified with a pseudo-Goldstone boson, the axion. At low-energies, the physical spectrum contains, besides the usual 2HDM degrees of freedom, the ultralight axion field, $a$, and a TeV-scale CP-even field. Moreover, we have shown that the imposition of orthogonality amongst the Goldstone bosons fixes the physical values for the $U(1)_{PQ}$ and $U(1)_X$ charges. As a direct consequence of this procedure, the tree-level mass mixing between the SM and the extra gauge boson, $Z^\prime$, vanishes identically.

The Yukawa sector bears similarities with the flipped (or type-Y) 2HDM in that the right-handed charged leptons and up-type quarks couple to the same Higgs doublet $\Phi_u$, while the down-type quarks and the neutral leptons couple to $\Phi_d$. To ensure gauge anomaly cancellation and generate small neutrino masses, we have introduced extra quarks, $k_{aL,R}$, chirally charged under $U(1)_X$, right-handed neutrinos, $\nu_{aR}$, and extra neutral leptons, $n_{aL,R}$.
The extra quarks get masses proportional to $v_\varphi$ and mix with the SM down-type quarks, leading to FCNCs mediated not only by $Z^\prime$ but also $a$, the axion field. For the neutral leptons, a Dirac seesaw mechanism generates small masses for the active neutrinos, see Fig. \ref{fig:DSS}. Neutrino mass suppression is controlled by the ratio $v_\varphi/v_\sigma$, where we have taken $v_\sigma=10^{12}$ GeV as the large PQ scale, suggesting that the $U(1)_X$ symmetry is broken at a much lower scale, $v_\varphi$. On the other hand, since $Z^\prime$ and the quarks $k_{a}$, get masses around $v_\varphi$, this scale is constrained from below by experiments and cosmology. These features suggest that $v_\varphi$ lies within a phenomenologically rich region, and we have assumed $v_\varphi= 10^4$ GeV as a benchmark.

Finally, we have studied the properties of the axion in Sec. \ref{sec:axion}. The axion has components along all the scalars but is mostly projected along the CP-odd component of the singlet $\sigma$ with a decay constant: $f_a\simeq v_\sigma/N_{DW}=v_\sigma/3$. For the chosen benchmark of $v_\sigma = 10^{12}$ GeV, which leads to naturally small neutrino masses, the associated axion may account for the totality of the observed cold dark matter, considering the pre-inflationary scenario for axion production. Despite the presence of extra fermions, the axion coupling to photons depends on the same anomaly coefficient ratio as in the type-II DFSZ model: $C_{a\gamma}/C_{ag}= E/N=2/3$. Fig. \ref{fig:anuplot} shows that the preferred region for neutrino masses and axion dark matter may be tested by forthcoming axion experiments looking for axion-photon interactions. Furthermore, we have derived the axion couplings to fermions and noticed that a rather weak lower bound for $f_a$ is obtained from flavour-violating processes, such as $K^+ \to \pi^+ a$, as a consequence of an additional suppression from $v_\varphi$.

\acknowledgments
A. G. Dias and D. S. V. Gon\c{c}alves thank Conselho Nacional de Desenvolvimento Cient\'{\i}fico e Tecnol\'ogico (CNPq) for financial support under the grants 305802/2019-4 and 130278/2020-3, respectively. J. Leite acknowledges financial support via grant 2017/23027-2, S\~ao Paulo Research Foundation (FAPESP).

\appendix

\section{Anomaly coefficients}
\label{app:anomalies}

The full Lagrangian of our model -- including the scalar potential, Eq. (\ref{V}), and the Yukawa interactions, Eq. (\ref{Yuk}) -- satisfies the $U(1)_{global}$ transformations defined by the five independent charges: $(q^\prime_{n_L}, q^\prime_{Q_L}, q^\prime_{\Phi_u}, q^\prime_{\varphi}, q^\prime_{\sigma})$, given in Table \ref{tab1}. Among the global symmetries, only those which lead to vanishing anomaly coefficients in Eq. (\ref{eq:anomalies}) can be safely promoted to local. To find the anomaly-free solutions, we calculate the coefficients using the charges in the $U(1)_{global}$ column and set them to zero, {\it i.e.}
\begin{equation}\label{eq:afree}
\begin{split}
    I\propto& \sum_{quarks} \left( q^\prime_L - q^\prime_R \right) = 3 q^\prime_\sigma =0 \,, \\
    II\propto& \sum_{L_L,Q_L} q^\prime_L = -12 q^\prime_{\sigma} +6\tilde{q}=0 \,, \\
    III\propto& \sum \left(Y_L^2 q^\prime_L - Y_R^2 q^\prime_R \right)= 4 q^\prime_{\sigma} - \frac{3}{2}\tilde{q}=0, \\
    IV\propto& \sum \left(Y_L q^{\prime 2}_L - Y_R q^{\prime 2}_R\right) = 3q^\prime_{\sigma}\left(6 q^\prime_{\Phi_u}-2 q^\prime_{Q_L}+q^\prime_{\sigma} \right)- 6 q^\prime_{\Phi_u} \tilde{q}=0 \,, \\
            V\propto& \sum \left(q^{\prime 3}_L -  q^{\prime 3}_R\right) = 9q^\prime_\sigma\left[7q^{\prime 2}_{\Phi_u}- (6 q^\prime_{\Phi_u}  - 3q^\prime_{Q_L})q^\prime_{Q_L} - q^{\prime 2}_{n_L} + (4 q^\prime_{n_L} -2 q^\prime_{\varphi}) q^\prime_{\varphi} \right] \,, \\
            & \quad\quad\quad\quad\quad\quad\quad\quad\!\!\!+ 9q^{\prime2}_\sigma\left( 3q^\prime_{\Phi_u} -3q^\prime_{Q_L} +3q^\prime_{n_L} -4q^\prime_{\varphi} \right) -12 q^{\prime3}_\sigma - 18 q^{\prime 2}_{\Phi_u}\tilde{q} =0\,, \\
    VI\propto&  \sum \left(q^\prime_L - q^\prime_R\right) = 6q^\prime_\sigma=0 \,,
\end{split}
\end{equation}
where $\tilde{q} = 3 q^\prime_{Q_L} + q^\prime_{n_L} - q^\prime_{\Phi_u} - q^\prime_{\varphi}$; the sum in $I$ takes only quarks into account, $II$ considers only fermion doublets, whereas for the remaining coefficients all fermions contribute. It is easy to see that the equations above are simultaneously satisfied when $q^\prime_\sigma=\tilde{q}=0$, reducing the number of free parameters from five to three. Finally, by renaming the three independent charges of this subset as $(l_{Q_L}, l_{\Phi_u}, l_\varphi)$, we obtain the $U(1)_{afree}$ column of Table \ref{tab1}.

\section{Diagonalisation}
\label{app:block-diagon}

Our task is to diagonalise the fermion mass matrices present in this work. To this end, we will first see how to block diagonalise a Hermitian matrix. Then, we will block diagonalise a non-Hermitian matrix as a generalisation of the previous procedure. Lastly, we will show how all the fermion mass matrices in this work can be fully diaogonalised.

\subsection{Block diagonalisation of Hermitian matrices}

Consider a $6 \times 6$ Hermitian matrix $M$. If we want to block diagonalise this matrix, it is sufficient to find a unitary matrix $R$, such that
\begin{equation}
    \label{eq:rmrd}
    R^{\dagger} M R = \text{diag}(M_1, M_2)\, ,
\end{equation}
where $M_i$, $i=1,2$, are $3 \times 3$ matrices. The procedure outlined here will work whenever the matrix $M$ can be written as
\begin{equation}
    M = \begin{pmatrix}
    m_1 A_1 & m_2 A_2 \\
    m_2 A_2^{\dagger} & m_3 A_3
    \end{pmatrix} \, ,
\end{equation}
where, $A_1$, $A_3$ are Hermitian matrices, the entries of the matrices $A_i$, $i=1,2,3$ are dimensionless and of order unity, whereas $m_i$ represent mass scales that satisfy the hierarchy $m_3 \gg m_1, m_2$. To achieve the block diagonalisation, we parameterise the unitary matrix $R$ as
\begin{equation}\label{Rmat}
    R = \begin{pmatrix}
                \sqrt{1-BB^{\dagger}} & B \\
                -B^{\dagger} & \sqrt{1-B^{\dagger}B}
            \end{pmatrix} \, ,
\end{equation}
where $B$ is a general complex $3 \times 3$ matrix to be determined. The square roots should be seen as series expansions in $B$ \cite{Grimus:2000vj,Hettmansperger:2011bt}
\begin{equation}
    \sqrt{1-BB^{\dagger}} = 1 - \frac{1}{2}BB^{\dagger} - \frac{1}{8} BB^{\dagger} BB^{\dagger} - ... \, .
\end{equation}
In turn, $B$ can also be expanded as $ B = B_1 + B_2 + B_3 + ... $, where $B_n$ is of order $\epsilon^n$ in the expansion parameter $\epsilon = m_2/m_3\ll1$. An alternative parameterisation for the matrix $R$ can be found in Ref. \cite{Korner:1992zk}.

\begin{equation}
    \label{eq:B}
    B \approx B_1 = \epsilon A_2 A_3^{-1} \, .
\end{equation}

\subsection{Block diagonalisation of non-Hermitian matrices}

Now, consider a $6 \times 6$ non-Hermitian matrix $M^F$. Our next task is to block diagonalise the down-type quark and the neutral lepton mass matrices. As neither of them is Hermitian, instead of the unitary transformation in Eq. (\ref{eq:rmrd}), a bi-unitary transformation is required to block diagonalise each of them,
\begin{equation}
    \label{eq:bi}
   R_L^{F\dagger} M^F R_R^F = \text{diag}(M^{F_1}, M^{F_2}) \equiv M^F_{block} \, ,
\end{equation}
where, $R_{L,R}^F$ are unitary matrices and $F_{1,2}$ are the components of the basis $F = D,N$. The problem is solved if we find the matrices $R_{L,R}^F$. One way of doing it is to break the bi-unitary transformation into two unitary transformations. Multiplying Eq. (\ref{eq:bi}) by its Hermitian conjugate gives
\begin{equation}
    \label{eq:uniL}
   M^F_{block} M_{block}^{F \dagger} = R_L^{F \dagger} M^F M^{F \dagger} R_L^F \equiv \mathcal{D}_L^F \, .
\end{equation}
Similarly, inverting the product order, we multiply the Hermitian conjugate by Eq. (\ref{eq:bi}) and find
\begin{equation}
    \label{eq:uniR}
     M_{block}^{F \dagger} M^F_{block}  = R_R^{F \dagger} M^{F \dagger} M^F R_R^F \equiv \mathcal{D}_R^F \, .
\end{equation}

The matrices $\mathcal{D}_{L,R}^F$ are block diagonal since $M^F_{block}$ also is. The matrices $M^F M^{F \dagger}$ and $M^{F\dagger} M^F$ are Hermitian, as a result, the Eqs. (\ref{eq:uniL}) and (\ref{eq:uniR}) represent unitary transformations, and we can make use of Eq. (\ref{Rmat}) to find the unitary matrices $R_{L,R}^F$ that block diagonalise $M^F$, up to the desired order. Thus, we can write the unitary matrices that diagonalise $M^D$ and $M^N$, in Eqs. (\ref{downMM}) and (\ref{ssmm}), respectively, in terms of the contributions in Table \ref{tab3}.
\begin{table}[h]
\begin{tabular}{|c|c|c|}
\hline
            & $D$        & $N$      \\\hline\hline
$B_L^F$     & $\frac{\mu v_d}{v_\varphi^2} y^d y^{\mu \dagger} \left( y^k y^{k \dagger} \right)^{-1}$   & $\frac{v_d}{v_\sigma} y^n y^{\beta \dagger} \left( y^\beta y^{\beta \dagger} \right)^{-1}$      \\\hline
$B_R^F$     & $\frac{\mu}{v_\varphi} y^{\mu \dagger} y^k \left( y^{k \dagger} y^k \right)^{-1}$  & $\frac{v_\varphi}{v_\sigma} y^{\alpha \dagger} y^\beta \left( y^{\beta \dagger} y^\beta \right)^{-1}$    \\\hline
\end{tabular}
\caption{Approximation for the $3 \times 3$ matrices $B_{L,R}^F$, with $F = D,N$, up to the first order term in the expansion. The orders of magnitude of $B_{L,R}^F$ are given by the vev relations for Yukawa matrices of order unity.}
\label{tab3}
\end{table}

\subsection{Diagonalisation of the mass matrices}

We are ready to diagonalise the mass matrices of all fermions present in this work. Consider a $6 \times 6$ block diagonal matrix, $M^F_{block}$, which is composed of two non-Hermitian $3 \times 3$ blocks, $M^{F_{1,2}}$. This is the structure for the down-type quarks and neutral leptons of the model after the block diagonalisation. To completely diagonalise $M^F_{block}$, we need another bi-unitary transformation,
\begin{equation}
    \label{eq:total-diag}
    \mathcal{V}_L^{F \dagger} M^F_{block} \mathcal{V}_R^F = M^{F^\prime} \, ,
\end{equation}
where $ M^{F^\prime}$ is the diagonal matrix for the mass eigenstates in the basis $F^\prime = D^\prime, N^\prime$, $\mathcal{V}_{L,R}^F = \text{diag}\left( V_{L,R}^{F_1}, V_{L,R}^{F_2} \right)$ are $6 \times 6$ unitary matrices, and $V_{L,R}^{F_{1,2}}$ are the unitary $3 \times 3$ matrices that diagonalise the upper and lower blocks of $\text{diag}(M^{F_1}, M^{F_2})$, {\it i.e.} $V_L^{F_i \dagger} M^{F_i} V_R^{F_i} = M^{F^\prime_i}$, $i=1,2$. Comparing Eqs. (\ref{eq:bi}) and (\ref{eq:total-diag}), we find
\begin{equation}
    \left( R_L^F \mathcal{V}_L^F \right)^{\dagger} M^F R_R^F \mathcal{V}_R^F= M^{F^\prime} \, ,
\end{equation}
that is, the combined effect of two diagonalisation steps is equivalent to a complete diagonalisation performed by the unitary matrices
\begin{equation}
    U_{L,R}^F \equiv R_{L,R}^F \mathcal{V}_{L,R}^F \, .
\end{equation}
Therefore, we can summarise all the individual diagonalisation procedures as
\begin{equation}
    U_L^{F \dagger} M^F U_R^F = M^{F^\prime} \,,
\end{equation}
where $F = e,N,u,D$, and $U_{R,L}^F$ has the same dimension as the corresponding mass matrix $M^F$.

\bibliographystyle{apsrev4-1}

\bibliography{myrefs}

\begin{thebibliography}{88}%
\makeatletter
\providecommand \@ifxundefined [1]{%
 \@ifx{#1\undefined}
}%
\providecommand \@ifnum [1]{%
 \ifnum #1\expandafter \@firstoftwo
 \else \expandafter \@secondoftwo
 \fi
}%
\providecommand \@ifx [1]{%
 \ifx #1\expandafter \@firstoftwo
 \else \expandafter \@secondoftwo
 \fi
}%
\providecommand \natexlab [1]{#1}%
\providecommand \enquote  [1]{``#1''}%
\providecommand \bibnamefont  [1]{#1}%
\providecommand \bibfnamefont [1]{#1}%
\providecommand \citenamefont [1]{#1}%
\providecommand \href@noop [0]{\@secondoftwo}%
\providecommand \href [0]{\begingroup \@sanitize@url \@href}%
\providecommand \@href[1]{\@@startlink{#1}\@@href}%
\providecommand \@@href[1]{\endgroup#1\@@endlink}%
\providecommand \@sanitize@url [0]{\catcode `\\12\catcode `\$12\catcode
  `\&12\catcode `\#12\catcode `\^12\catcode `\_12\catcode `\%12\relax}%
\providecommand \@@startlink[1]{}%
\providecommand \@@endlink[0]{}%
\providecommand \url  [0]{\begingroup\@sanitize@url \@url }%
\providecommand \@url [1]{\endgroup\@href {#1}{\urlprefix }}%
\providecommand \urlprefix  [0]{URL }%
\providecommand \Eprint [0]{\href }%
\providecommand \doibase [0]{http://dx.doi.org/}%
\providecommand \selectlanguage [0]{\@gobble}%
\providecommand \bibinfo  [0]{\@secondoftwo}%
\providecommand \bibfield  [0]{\@secondoftwo}%
\providecommand \translation [1]{[#1]}%
\providecommand \BibitemOpen [0]{}%
\providecommand \bibitemStop [0]{}%
\providecommand \bibitemNoStop [0]{.\EOS\space}%
\providecommand \EOS [0]{\spacefactor3000\relax}%
\providecommand \BibitemShut  [1]{\csname bibitem#1\endcsname}%
\let\auto@bib@innerbib\@empty
\bibitem [{\citenamefont {Fukuda}\ \emph {et~al.}(1998)\citenamefont {Fukuda}
  \emph {et~al.}}]{Fukuda:1998mi}%
  \BibitemOpen
  \bibfield  {author} {\bibinfo {author} {\bibfnamefont {Y.}~\bibnamefont
  {Fukuda}} \emph {et~al.} (\bibinfo {collaboration} {Super-Kamiokande}),\
  }\href {\doibase 10.1103/PhysRevLett.81.1562} {\bibfield  {journal} {\bibinfo
   {journal} {Phys. Rev. Lett.}\ }\textbf {\bibinfo {volume} {81}},\ \bibinfo
  {pages} {1562} (\bibinfo {year} {1998})},\ \Eprint
  {http://arxiv.org/abs/hep-ex/9807003} {arXiv:hep-ex/9807003} \BibitemShut
  {NoStop}%
\bibitem [{\citenamefont {Ahmad}\ \emph {et~al.}(2001)\citenamefont {Ahmad}
  \emph {et~al.}}]{Ahmad:2001an}%
  \BibitemOpen
  \bibfield  {author} {\bibinfo {author} {\bibfnamefont {Q.~R.}\ \bibnamefont
  {Ahmad}} \emph {et~al.} (\bibinfo {collaboration} {SNO}),\ }\href {\doibase
  10.1103/PhysRevLett.87.071301} {\bibfield  {journal} {\bibinfo  {journal}
  {Phys. Rev. Lett.}\ }\textbf {\bibinfo {volume} {87}},\ \bibinfo {pages}
  {071301} (\bibinfo {year} {2001})},\ \Eprint
  {http://arxiv.org/abs/nucl-ex/0106015} {arXiv:nucl-ex/0106015} \BibitemShut
  {NoStop}%
\bibitem [{\citenamefont {Ahmad}\ \emph {et~al.}(2002)\citenamefont {Ahmad}
  \emph {et~al.}}]{Ahmad:2002jz}%
  \BibitemOpen
  \bibfield  {author} {\bibinfo {author} {\bibfnamefont {Q.~R.}\ \bibnamefont
  {Ahmad}} \emph {et~al.} (\bibinfo {collaboration} {SNO}),\ }\href {\doibase
  10.1103/PhysRevLett.89.011301} {\bibfield  {journal} {\bibinfo  {journal}
  {Phys. Rev. Lett.}\ }\textbf {\bibinfo {volume} {89}},\ \bibinfo {pages}
  {011301} (\bibinfo {year} {2002})},\ \Eprint
  {http://arxiv.org/abs/nucl-ex/0204008} {arXiv:nucl-ex/0204008} \BibitemShut
  {NoStop}%
\bibitem [{\citenamefont {Valle}\ and\ \citenamefont
  {Romao}(2015)}]{Valle:2015pba}%
  \BibitemOpen
  \bibfield  {author} {\bibinfo {author} {\bibfnamefont {J.~W.~F.}\
  \bibnamefont {Valle}}\ and\ \bibinfo {author} {\bibfnamefont {J.~C.}\
  \bibnamefont {Romao}},\ }\href@noop {} {\emph {\bibinfo {title} {{Neutrinos
  in high energy and astroparticle physics}}}},\ Physics textbook\ (\bibinfo
  {publisher} {Wiley-VCH},\ \bibinfo {address} {Weinheim},\ \bibinfo {year}
  {2015})\BibitemShut {NoStop}%
\bibitem [{\citenamefont {Zee}(1986)}]{Zee:1985id}%
  \BibitemOpen
  \bibfield  {author} {\bibinfo {author} {\bibfnamefont {A.}~\bibnamefont
  {Zee}},\ }\href {\doibase 10.1016/0550-3213(86)90475-X} {\bibfield  {journal}
  {\bibinfo  {journal} {Nucl. Phys. B}\ }\textbf {\bibinfo {volume} {264}},\
  \bibinfo {pages} {99} (\bibinfo {year} {1986})}\BibitemShut {NoStop}%
\bibitem [{\citenamefont {Babu}(1988)}]{Babu:1988ki}%
  \BibitemOpen
  \bibfield  {author} {\bibinfo {author} {\bibfnamefont {K.~S.}\ \bibnamefont
  {Babu}},\ }\href {\doibase 10.1016/0370-2693(88)91584-5} {\bibfield
  {journal} {\bibinfo  {journal} {Phys. Lett. B}\ }\textbf {\bibinfo {volume}
  {203}},\ \bibinfo {pages} {132} (\bibinfo {year} {1988})}\BibitemShut
  {NoStop}%
\bibitem [{\citenamefont {Pilaftsis}(1992)}]{Pilaftsis:1991ug}%
  \BibitemOpen
  \bibfield  {author} {\bibinfo {author} {\bibfnamefont {A.}~\bibnamefont
  {Pilaftsis}},\ }\href {\doibase 10.1007/BF01482590} {\bibfield  {journal}
  {\bibinfo  {journal} {Z. Phys. C}\ }\textbf {\bibinfo {volume} {55}},\
  \bibinfo {pages} {275} (\bibinfo {year} {1992})},\ \Eprint
  {http://arxiv.org/abs/hep-ph/9901206} {arXiv:hep-ph/9901206} \BibitemShut
  {NoStop}%
\bibitem [{\citenamefont {Ma}(2006)}]{Ma:2006km}%
  \BibitemOpen
  \bibfield  {author} {\bibinfo {author} {\bibfnamefont {E.}~\bibnamefont
  {Ma}},\ }\href {\doibase 10.1103/PhysRevD.73.077301} {\bibfield  {journal}
  {\bibinfo  {journal} {Phys. Rev. D}\ }\textbf {\bibinfo {volume} {73}},\
  \bibinfo {pages} {077301} (\bibinfo {year} {2006})},\ \Eprint
  {http://arxiv.org/abs/hep-ph/0601225} {arXiv:hep-ph/0601225} \BibitemShut
  {NoStop}%
\bibitem [{\citenamefont {de~Salas}\ \emph {et~al.}(2021)\citenamefont
  {de~Salas}, \citenamefont {Forero}, \citenamefont {Gariazzo}, \citenamefont
  {Mart\'\i{}nez-Mirav\'e}, \citenamefont {Mena}, \citenamefont {Ternes},
  \citenamefont {T\'ortola},\ and\ \citenamefont {Valle}}]{deSalas:2020pgw}%
  \BibitemOpen
  \bibfield  {author} {\bibinfo {author} {\bibfnamefont {P.~F.}\ \bibnamefont
  {de~Salas}}, \bibinfo {author} {\bibfnamefont {D.~V.}\ \bibnamefont
  {Forero}}, \bibinfo {author} {\bibfnamefont {S.}~\bibnamefont {Gariazzo}},
  \bibinfo {author} {\bibfnamefont {P.}~\bibnamefont {Mart\'\i{}nez-Mirav\'e}},
  \bibinfo {author} {\bibfnamefont {O.}~\bibnamefont {Mena}}, \bibinfo {author}
  {\bibfnamefont {C.~A.}\ \bibnamefont {Ternes}}, \bibinfo {author}
  {\bibfnamefont {M.}~\bibnamefont {T\'ortola}}, \ and\ \bibinfo {author}
  {\bibfnamefont {J.~W.~F.}\ \bibnamefont {Valle}},\ }\href {\doibase
  10.1007/JHEP02(2021)071} {\bibfield  {journal} {\bibinfo  {journal} {JHEP}\
  }\textbf {\bibinfo {volume} {02}},\ \bibinfo {pages} {071} (\bibinfo {year}
  {2021})},\ \Eprint {http://arxiv.org/abs/2006.11237} {arXiv:2006.11237
  [hep-ph]} \BibitemShut {NoStop}%
\bibitem [{\citenamefont {Aghanim}\ \emph {et~al.}(2020)\citenamefont {Aghanim}
  \emph {et~al.}}]{Aghanim:2018eyx}%
  \BibitemOpen
  \bibfield  {author} {\bibinfo {author} {\bibfnamefont {N.}~\bibnamefont
  {Aghanim}} \emph {et~al.} (\bibinfo {collaboration} {Planck}),\ }\href
  {\doibase 10.1051/0004-6361/201833910} {\bibfield  {journal} {\bibinfo
  {journal} {Astron. Astrophys.}\ }\textbf {\bibinfo {volume} {641}},\ \bibinfo
  {pages} {A6} (\bibinfo {year} {2020})},\ \Eprint
  {http://arxiv.org/abs/1807.06209} {arXiv:1807.06209 [astro-ph.CO]}
  \BibitemShut {NoStop}%
\bibitem [{\citenamefont {Bertone}\ and\ \citenamefont
  {Hooper}(2018)}]{Bertone:2016nfn}%
  \BibitemOpen
  \bibfield  {author} {\bibinfo {author} {\bibfnamefont {G.}~\bibnamefont
  {Bertone}}\ and\ \bibinfo {author} {\bibfnamefont {D.}~\bibnamefont
  {Hooper}},\ }\href {\doibase 10.1103/RevModPhys.90.045002} {\bibfield
  {journal} {\bibinfo  {journal} {Rev. Mod. Phys.}\ }\textbf {\bibinfo {volume}
  {90}},\ \bibinfo {pages} {045002} (\bibinfo {year} {2018})},\ \Eprint
  {http://arxiv.org/abs/1605.04909} {arXiv:1605.04909 [astro-ph.CO]}
  \BibitemShut {NoStop}%
\bibitem [{\citenamefont {Aprile}\ \emph {et~al.}(2018)\citenamefont {Aprile}
  \emph {et~al.}}]{Aprile:2018dbl}%
  \BibitemOpen
  \bibfield  {author} {\bibinfo {author} {\bibfnamefont {E.}~\bibnamefont
  {Aprile}} \emph {et~al.} (\bibinfo {collaboration} {XENON}),\ }\href
  {\doibase 10.1103/PhysRevLett.121.111302} {\bibfield  {journal} {\bibinfo
  {journal} {Phys. Rev. Lett.}\ }\textbf {\bibinfo {volume} {121}},\ \bibinfo
  {pages} {111302} (\bibinfo {year} {2018})},\ \Eprint
  {http://arxiv.org/abs/1805.12562} {arXiv:1805.12562 [astro-ph.CO]}
  \BibitemShut {NoStop}%
\bibitem [{\citenamefont {Peccei}\ and\ \citenamefont
  {Quinn}(1977)}]{Peccei:1977hh}%
  \BibitemOpen
  \bibfield  {author} {\bibinfo {author} {\bibfnamefont {R.~D.}\ \bibnamefont
  {Peccei}}\ and\ \bibinfo {author} {\bibfnamefont {H.~R.}\ \bibnamefont
  {Quinn}},\ }\href {\doibase 10.1103/PhysRevLett.38.1440} {\bibfield
  {journal} {\bibinfo  {journal} {Phys. Rev. Lett.}\ }\textbf {\bibinfo
  {volume} {38}},\ \bibinfo {pages} {1440} (\bibinfo {year} {1977})},\ \bibinfo
  {note} {[,328(1977)]}\BibitemShut {NoStop}%
\bibitem [{\citenamefont {Weinberg}(1978)}]{Weinberg:1977ma}%
  \BibitemOpen
  \bibfield  {author} {\bibinfo {author} {\bibfnamefont {S.}~\bibnamefont
  {Weinberg}},\ }\href {\doibase 10.1103/PhysRevLett.40.223} {\bibfield
  {journal} {\bibinfo  {journal} {Phys. Rev. Lett.}\ }\textbf {\bibinfo
  {volume} {40}},\ \bibinfo {pages} {223} (\bibinfo {year} {1978})}\BibitemShut
  {NoStop}%
\bibitem [{\citenamefont {Wilczek}(1978)}]{Wilczek:1977pj}%
  \BibitemOpen
  \bibfield  {author} {\bibinfo {author} {\bibfnamefont {F.}~\bibnamefont
  {Wilczek}},\ }\href {\doibase 10.1103/PhysRevLett.40.279} {\bibfield
  {journal} {\bibinfo  {journal} {Phys. Rev. Lett.}\ }\textbf {\bibinfo
  {volume} {40}},\ \bibinfo {pages} {279} (\bibinfo {year} {1978})}\BibitemShut
  {NoStop}%
\bibitem [{\citenamefont {Kim}\ and\ \citenamefont
  {Carosi}(2010)}]{Kim:2008hd}%
  \BibitemOpen
  \bibfield  {author} {\bibinfo {author} {\bibfnamefont {J.~E.}\ \bibnamefont
  {Kim}}\ and\ \bibinfo {author} {\bibfnamefont {G.}~\bibnamefont {Carosi}},\
  }\href {\doibase 10.1103/RevModPhys.82.557} {\bibfield  {journal} {\bibinfo
  {journal} {Rev. Mod. Phys.}\ }\textbf {\bibinfo {volume} {82}},\ \bibinfo
  {pages} {557} (\bibinfo {year} {2010})},\ \bibinfo {note} {[Erratum:
  Rev.Mod.Phys. 91, 049902 (2019)]},\ \Eprint {http://arxiv.org/abs/0807.3125}
  {arXiv:0807.3125 [hep-ph]} \BibitemShut {NoStop}%
\bibitem [{\citenamefont {Hook}(2019)}]{Hook:2018dlk}%
  \BibitemOpen
  \bibfield  {author} {\bibinfo {author} {\bibfnamefont {A.}~\bibnamefont
  {Hook}},\ }\href@noop {} {\bibfield  {journal} {\bibinfo  {journal} {PoS}\
  }\textbf {\bibinfo {volume} {TASI2018}},\ \bibinfo {pages} {004} (\bibinfo
  {year} {2019})},\ \Eprint {http://arxiv.org/abs/1812.02669} {arXiv:1812.02669
  [hep-ph]} \BibitemShut {NoStop}%
\bibitem [{\citenamefont {Abbott}\ and\ \citenamefont
  {Sikivie}(1983)}]{Abbott:1982af}%
  \BibitemOpen
  \bibfield  {author} {\bibinfo {author} {\bibfnamefont {L.}~\bibnamefont
  {Abbott}}\ and\ \bibinfo {author} {\bibfnamefont {P.}~\bibnamefont
  {Sikivie}},\ }\href {\doibase 10.1016/0370-2693(83)90638-X} {\bibfield
  {journal} {\bibinfo  {journal} {Phys.Lett.}\ }\textbf {\bibinfo {volume}
  {B120}},\ \bibinfo {pages} {133} (\bibinfo {year} {1983})}\BibitemShut
  {NoStop}%
\bibitem [{\citenamefont {Preskill}\ \emph {et~al.}(1983)\citenamefont
  {Preskill}, \citenamefont {Wise},\ and\ \citenamefont
  {Wilczek}}]{Preskill:1982cy}%
  \BibitemOpen
  \bibfield  {author} {\bibinfo {author} {\bibfnamefont {J.}~\bibnamefont
  {Preskill}}, \bibinfo {author} {\bibfnamefont {M.~B.}\ \bibnamefont {Wise}},
  \ and\ \bibinfo {author} {\bibfnamefont {F.}~\bibnamefont {Wilczek}},\ }\href
  {\doibase 10.1016/0370-2693(83)90637-8} {\bibfield  {journal} {\bibinfo
  {journal} {Phys.Lett.}\ }\textbf {\bibinfo {volume} {B120}},\ \bibinfo
  {pages} {127} (\bibinfo {year} {1983})}\BibitemShut {NoStop}%
\bibitem [{\citenamefont {Dine}\ and\ \citenamefont
  {Fischler}(1983)}]{Dine:1982ah}%
  \BibitemOpen
  \bibfield  {author} {\bibinfo {author} {\bibfnamefont {M.}~\bibnamefont
  {Dine}}\ and\ \bibinfo {author} {\bibfnamefont {W.}~\bibnamefont
  {Fischler}},\ }\href {\doibase 10.1016/0370-2693(83)90639-1} {\bibfield
  {journal} {\bibinfo  {journal} {Phys.Lett.}\ }\textbf {\bibinfo {volume}
  {B120}},\ \bibinfo {pages} {137} (\bibinfo {year} {1983})}\BibitemShut
  {NoStop}%
\bibitem [{\citenamefont {Langacker}\ \emph {et~al.}(1986)\citenamefont
  {Langacker}, \citenamefont {Peccei},\ and\ \citenamefont
  {Yanagida}}]{Langacker:1986rj}%
  \BibitemOpen
  \bibfield  {author} {\bibinfo {author} {\bibfnamefont {P.}~\bibnamefont
  {Langacker}}, \bibinfo {author} {\bibfnamefont {R.~D.}\ \bibnamefont
  {Peccei}}, \ and\ \bibinfo {author} {\bibfnamefont {T.}~\bibnamefont
  {Yanagida}},\ }\href {\doibase 10.1142/S0217732386000683} {\bibfield
  {journal} {\bibinfo  {journal} {Mod. Phys. Lett. A}\ }\textbf {\bibinfo
  {volume} {1}},\ \bibinfo {pages} {541} (\bibinfo {year} {1986})}\BibitemShut
  {NoStop}%
\bibitem [{\citenamefont {Mohapatra}\ and\ \citenamefont
  {Senjanovic}(1983)}]{Mohapatra:1982tc}%
  \BibitemOpen
  \bibfield  {author} {\bibinfo {author} {\bibfnamefont {R.~N.}\ \bibnamefont
  {Mohapatra}}\ and\ \bibinfo {author} {\bibfnamefont {G.}~\bibnamefont
  {Senjanovic}},\ }\href {\doibase 10.1007/BF01577819} {\bibfield  {journal}
  {\bibinfo  {journal} {Z. Phys. C}\ }\textbf {\bibinfo {volume} {17}},\
  \bibinfo {pages} {53} (\bibinfo {year} {1983})}\BibitemShut {NoStop}%
\bibitem [{\citenamefont {Dias}\ \emph {et~al.}(2014)\citenamefont {Dias},
  \citenamefont {Machado}, \citenamefont {Nishi}, \citenamefont {Ringwald},\
  and\ \citenamefont {Vaudrevange}}]{Dias:2014osa}%
  \BibitemOpen
  \bibfield  {author} {\bibinfo {author} {\bibfnamefont {A.}~\bibnamefont
  {Dias}}, \bibinfo {author} {\bibfnamefont {A.}~\bibnamefont {Machado}},
  \bibinfo {author} {\bibfnamefont {C.}~\bibnamefont {Nishi}}, \bibinfo
  {author} {\bibfnamefont {A.}~\bibnamefont {Ringwald}}, \ and\ \bibinfo
  {author} {\bibfnamefont {P.}~\bibnamefont {Vaudrevange}},\ }\href {\doibase
  10.1007/JHEP06(2014)037} {\bibfield  {journal} {\bibinfo  {journal} {JHEP}\
  }\textbf {\bibinfo {volume} {1406}},\ \bibinfo {pages} {037} (\bibinfo {year}
  {2014})},\ \Eprint {http://arxiv.org/abs/1403.5760} {arXiv:1403.5760
  [hep-ph]} \BibitemShut {NoStop}%
\bibitem [{\citenamefont {Bertolini}\ \emph {et~al.}(2015)\citenamefont
  {Bertolini}, \citenamefont {Di~Luzio}, \citenamefont {Kole{\v{s}}ov{\'a}},\
  and\ \citenamefont {Malinsk{\'y}}}]{Bertolini:2014aia}%
  \BibitemOpen
  \bibfield  {author} {\bibinfo {author} {\bibfnamefont {S.}~\bibnamefont
  {Bertolini}}, \bibinfo {author} {\bibfnamefont {L.}~\bibnamefont {Di~Luzio}},
  \bibinfo {author} {\bibfnamefont {H.}~\bibnamefont {Kole{\v{s}}ov{\'a}}}, \
  and\ \bibinfo {author} {\bibfnamefont {M.}~\bibnamefont {Malinsk{\'y}}},\
  }\href {\doibase 10.1103/PhysRevD.91.055014} {\bibfield  {journal} {\bibinfo
  {journal} {Phys.Rev.}\ }\textbf {\bibinfo {volume} {D91}},\ \bibinfo {pages}
  {055014} (\bibinfo {year} {2015})},\ \Eprint {http://arxiv.org/abs/1412.7105}
  {arXiv:1412.7105 [hep-ph]} \BibitemShut {NoStop}%
\bibitem [{\citenamefont {Clarke}\ and\ \citenamefont
  {Volkas}(2016)}]{Clarke:2015bea}%
  \BibitemOpen
  \bibfield  {author} {\bibinfo {author} {\bibfnamefont {J.~D.}\ \bibnamefont
  {Clarke}}\ and\ \bibinfo {author} {\bibfnamefont {R.~R.}\ \bibnamefont
  {Volkas}},\ }\href {\doibase 10.1103/PhysRevD.93.035001} {\bibfield
  {journal} {\bibinfo  {journal} {Phys. Rev. D}\ }\textbf {\bibinfo {volume}
  {93}},\ \bibinfo {pages} {035001} (\bibinfo {year} {2016})},\ \Eprint
  {http://arxiv.org/abs/1509.07243} {arXiv:1509.07243 [hep-ph]} \BibitemShut
  {NoStop}%
\bibitem [{\citenamefont {Ahn}\ and\ \citenamefont {Chun}(2016)}]{Ahn:2015pia}%
  \BibitemOpen
  \bibfield  {author} {\bibinfo {author} {\bibfnamefont {Y.}~\bibnamefont
  {Ahn}}\ and\ \bibinfo {author} {\bibfnamefont {E.~J.}\ \bibnamefont {Chun}},\
  }\href {\doibase 10.1016/j.physletb.2015.11.067} {\bibfield  {journal}
  {\bibinfo  {journal} {Phys.Lett.}\ }\textbf {\bibinfo {volume} {B752}},\
  \bibinfo {pages} {333} (\bibinfo {year} {2016})},\ \Eprint
  {http://arxiv.org/abs/1510.01015} {arXiv:1510.01015 [hep-ph]} \BibitemShut
  {NoStop}%
\bibitem [{\citenamefont {Ballesteros}\ \emph
  {et~al.}(2017{\natexlab{a}})\citenamefont {Ballesteros}, \citenamefont
  {Redondo}, \citenamefont {Ringwald},\ and\ \citenamefont
  {Tamarit}}]{Ballesteros:2016euj}%
  \BibitemOpen
  \bibfield  {author} {\bibinfo {author} {\bibfnamefont {G.}~\bibnamefont
  {Ballesteros}}, \bibinfo {author} {\bibfnamefont {J.}~\bibnamefont
  {Redondo}}, \bibinfo {author} {\bibfnamefont {A.}~\bibnamefont {Ringwald}}, \
  and\ \bibinfo {author} {\bibfnamefont {C.}~\bibnamefont {Tamarit}},\ }\href
  {\doibase 10.1103/PhysRevLett.118.071802} {\bibfield  {journal} {\bibinfo
  {journal} {Phys. Rev. Lett.}\ }\textbf {\bibinfo {volume} {118}},\ \bibinfo
  {pages} {071802} (\bibinfo {year} {2017}{\natexlab{a}})},\ \Eprint
  {http://arxiv.org/abs/1608.05414} {arXiv:1608.05414 [hep-ph]} \BibitemShut
  {NoStop}%
\bibitem [{\citenamefont {Ballesteros}\ \emph
  {et~al.}(2017{\natexlab{b}})\citenamefont {Ballesteros}, \citenamefont
  {Redondo}, \citenamefont {Ringwald},\ and\ \citenamefont
  {Tamarit}}]{Ballesteros:2016xej}%
  \BibitemOpen
  \bibfield  {author} {\bibinfo {author} {\bibfnamefont {G.}~\bibnamefont
  {Ballesteros}}, \bibinfo {author} {\bibfnamefont {J.}~\bibnamefont
  {Redondo}}, \bibinfo {author} {\bibfnamefont {A.}~\bibnamefont {Ringwald}}, \
  and\ \bibinfo {author} {\bibfnamefont {C.}~\bibnamefont {Tamarit}},\ }\href
  {\doibase 10.1088/1475-7516/2017/08/001} {\bibfield  {journal} {\bibinfo
  {journal} {JCAP}\ }\textbf {\bibinfo {volume} {08}},\ \bibinfo {pages} {001}
  (\bibinfo {year} {2017}{\natexlab{b}})},\ \Eprint
  {http://arxiv.org/abs/1610.01639} {arXiv:1610.01639 [hep-ph]} \BibitemShut
  {NoStop}%
\bibitem [{\citenamefont {Chen}\ and\ \citenamefont
  {Tsai}(2013)}]{Chen:2012baa}%
  \BibitemOpen
  \bibfield  {author} {\bibinfo {author} {\bibfnamefont {C.-S.}\ \bibnamefont
  {Chen}}\ and\ \bibinfo {author} {\bibfnamefont {L.-H.}\ \bibnamefont
  {Tsai}},\ }\href {\doibase 10.1103/PhysRevD.88.055015} {\bibfield  {journal}
  {\bibinfo  {journal} {Phys.Rev.}\ }\textbf {\bibinfo {volume} {D88}},\
  \bibinfo {pages} {055015} (\bibinfo {year} {2013})},\ \Eprint
  {http://arxiv.org/abs/1210.6264} {arXiv:1210.6264 [hep-ph]} \BibitemShut
  {NoStop}%
\bibitem [{\citenamefont {Gu}(2016)}]{Gu:2016hxh}%
  \BibitemOpen
  \bibfield  {author} {\bibinfo {author} {\bibfnamefont {P.-H.}\ \bibnamefont
  {Gu}},\ }\href {\doibase 10.1088/1475-7516/2016/07/004} {\bibfield  {journal}
  {\bibinfo  {journal} {JCAP}\ }\textbf {\bibinfo {volume} {1607}},\ \bibinfo
  {pages} {004} (\bibinfo {year} {2016})},\ \Eprint
  {http://arxiv.org/abs/1603.05070} {arXiv:1603.05070 [hep-ph]} \BibitemShut
  {NoStop}%
\bibitem [{\citenamefont {Peinado}\ \emph {et~al.}(2020)\citenamefont
  {Peinado}, \citenamefont {Reig}, \citenamefont {Srivastava},\ and\
  \citenamefont {Valle}}]{Peinado:2019mrn}%
  \BibitemOpen
  \bibfield  {author} {\bibinfo {author} {\bibfnamefont {E.}~\bibnamefont
  {Peinado}}, \bibinfo {author} {\bibfnamefont {M.}~\bibnamefont {Reig}},
  \bibinfo {author} {\bibfnamefont {R.}~\bibnamefont {Srivastava}}, \ and\
  \bibinfo {author} {\bibfnamefont {J.~W.~F.}\ \bibnamefont {Valle}},\ }\href
  {\doibase 10.1142/S021773232050176X} {\bibfield  {journal} {\bibinfo
  {journal} {Mod. Phys. Lett. A}\ }\textbf {\bibinfo {volume} {35}},\ \bibinfo
  {pages} {2050176} (\bibinfo {year} {2020})},\ \Eprint
  {http://arxiv.org/abs/1910.02961} {arXiv:1910.02961 [hep-ph]} \BibitemShut
  {NoStop}%
\bibitem [{\citenamefont {Baek}(2020)}]{Baek:2019wdn}%
  \BibitemOpen
  \bibfield  {author} {\bibinfo {author} {\bibfnamefont {S.}~\bibnamefont
  {Baek}},\ }\href {\doibase 10.1016/j.physletb.2020.135415} {\bibfield
  {journal} {\bibinfo  {journal} {Phys. Lett. B}\ }\textbf {\bibinfo {volume}
  {805}},\ \bibinfo {pages} {135415} (\bibinfo {year} {2020})},\ \Eprint
  {http://arxiv.org/abs/1911.04210} {arXiv:1911.04210 [hep-ph]} \BibitemShut
  {NoStop}%
\bibitem [{\citenamefont {Dias}\ \emph {et~al.}(2020)\citenamefont {Dias},
  \citenamefont {Leite}, \citenamefont {Valle},\ and\ \citenamefont
  {Vaquera-Araujo}}]{Dias:2020kbj}%
  \BibitemOpen
  \bibfield  {author} {\bibinfo {author} {\bibfnamefont {A.~G.}\ \bibnamefont
  {Dias}}, \bibinfo {author} {\bibfnamefont {J.}~\bibnamefont {Leite}},
  \bibinfo {author} {\bibfnamefont {J.~W.~F.}\ \bibnamefont {Valle}}, \ and\
  \bibinfo {author} {\bibfnamefont {C.~A.}\ \bibnamefont {Vaquera-Araujo}},\
  }\href {\doibase 10.1016/j.physletb.2020.135829} {\bibfield  {journal}
  {\bibinfo  {journal} {Phys. Lett. B}\ }\textbf {\bibinfo {volume} {810}},\
  \bibinfo {pages} {135829} (\bibinfo {year} {2020})},\ \Eprint
  {http://arxiv.org/abs/2008.10650} {arXiv:2008.10650 [hep-ph]} \BibitemShut
  {NoStop}%
\bibitem [{\citenamefont {Gando}\ \emph {et~al.}(2016)\citenamefont {Gando}
  \emph {et~al.}}]{KamLAND-Zen:2016pfg}%
  \BibitemOpen
  \bibfield  {author} {\bibinfo {author} {\bibfnamefont {A.}~\bibnamefont
  {Gando}} \emph {et~al.} (\bibinfo {collaboration} {KamLAND-Zen}),\ }\href
  {\doibase 10.1103/PhysRevLett.117.082503} {\bibfield  {journal} {\bibinfo
  {journal} {Phys. Rev. Lett.}\ }\textbf {\bibinfo {volume} {117}},\ \bibinfo
  {pages} {082503} (\bibinfo {year} {2016})},\ \bibinfo {note} {[Addendum:
  Phys.Rev.Lett. 117, 109903 (2016)]},\ \Eprint
  {http://arxiv.org/abs/1605.02889} {arXiv:1605.02889 [hep-ex]} \BibitemShut
  {NoStop}%
\bibitem [{\citenamefont {Branco}\ \emph {et~al.}(2012)\citenamefont {Branco},
  \citenamefont {Ferreira}, \citenamefont {Lavoura}, \citenamefont {Rebelo},
  \citenamefont {Sher},\ and\ \citenamefont {Silva}}]{Branco:2011iw}%
  \BibitemOpen
  \bibfield  {author} {\bibinfo {author} {\bibfnamefont {G.~C.}\ \bibnamefont
  {Branco}}, \bibinfo {author} {\bibfnamefont {P.~M.}\ \bibnamefont
  {Ferreira}}, \bibinfo {author} {\bibfnamefont {L.}~\bibnamefont {Lavoura}},
  \bibinfo {author} {\bibfnamefont {M.~N.}\ \bibnamefont {Rebelo}}, \bibinfo
  {author} {\bibfnamefont {M.}~\bibnamefont {Sher}}, \ and\ \bibinfo {author}
  {\bibfnamefont {J.~P.}\ \bibnamefont {Silva}},\ }\href {\doibase
  10.1016/j.physrep.2012.02.002} {\bibfield  {journal} {\bibinfo  {journal}
  {Phys. Rept.}\ }\textbf {\bibinfo {volume} {516}},\ \bibinfo {pages} {1}
  (\bibinfo {year} {2012})},\ \Eprint {http://arxiv.org/abs/1106.0034}
  {arXiv:1106.0034 [hep-ph]} \BibitemShut {NoStop}%
\bibitem [{\citenamefont {Ko}\ \emph {et~al.}(2012)\citenamefont {Ko},
  \citenamefont {Omura},\ and\ \citenamefont {Yu}}]{Ko:2012hd}%
  \BibitemOpen
  \bibfield  {author} {\bibinfo {author} {\bibfnamefont {P.}~\bibnamefont
  {Ko}}, \bibinfo {author} {\bibfnamefont {Y.}~\bibnamefont {Omura}}, \ and\
  \bibinfo {author} {\bibfnamefont {C.}~\bibnamefont {Yu}},\ }\href {\doibase
  10.1016/j.physletb.2012.09.019} {\bibfield  {journal} {\bibinfo  {journal}
  {Phys. Lett. B}\ }\textbf {\bibinfo {volume} {717}},\ \bibinfo {pages} {202}
  (\bibinfo {year} {2012})},\ \Eprint {http://arxiv.org/abs/1204.4588}
  {arXiv:1204.4588 [hep-ph]} \BibitemShut {NoStop}%
\bibitem [{\citenamefont {Ko}\ \emph {et~al.}(2014{\natexlab{a}})\citenamefont
  {Ko}, \citenamefont {Omura},\ and\ \citenamefont {Yu}}]{Ko:2014uka}%
  \BibitemOpen
  \bibfield  {author} {\bibinfo {author} {\bibfnamefont {P.}~\bibnamefont
  {Ko}}, \bibinfo {author} {\bibfnamefont {Y.}~\bibnamefont {Omura}}, \ and\
  \bibinfo {author} {\bibfnamefont {C.}~\bibnamefont {Yu}},\ }\href {\doibase
  10.1007/JHEP11(2014)054} {\bibfield  {journal} {\bibinfo  {journal} {JHEP}\
  }\textbf {\bibinfo {volume} {11}},\ \bibinfo {pages} {054} (\bibinfo {year}
  {2014}{\natexlab{a}})},\ \Eprint {http://arxiv.org/abs/1405.2138}
  {arXiv:1405.2138 [hep-ph]} \BibitemShut {NoStop}%
\bibitem [{\citenamefont {Okada}\ \emph {et~al.}(2020)\citenamefont {Okada},
  \citenamefont {Raut},\ and\ \citenamefont {Shafi}}]{Okada:2020cvq}%
  \BibitemOpen
  \bibfield  {author} {\bibinfo {author} {\bibfnamefont {N.}~\bibnamefont
  {Okada}}, \bibinfo {author} {\bibfnamefont {D.}~\bibnamefont {Raut}}, \ and\
  \bibinfo {author} {\bibfnamefont {Q.}~\bibnamefont {Shafi}},\ }\href
  {\doibase 10.1140/epjc/s10052-020-8343-6} {\bibfield  {journal} {\bibinfo
  {journal} {Eur. Phys. J. C}\ }\textbf {\bibinfo {volume} {80}},\ \bibinfo
  {pages} {1056} (\bibinfo {year} {2020})},\ \Eprint
  {http://arxiv.org/abs/2002.07110} {arXiv:2002.07110 [hep-ph]} \BibitemShut
  {NoStop}%
\bibitem [{\citenamefont {Campos}\ \emph {et~al.}(2017)\citenamefont {Campos},
  \citenamefont {Cogollo}, \citenamefont {Lindner}, \citenamefont {Melo},
  \citenamefont {Queiroz},\ and\ \citenamefont {Rodejohann}}]{Campos:2017dgc}%
  \BibitemOpen
  \bibfield  {author} {\bibinfo {author} {\bibfnamefont {M.~D.}\ \bibnamefont
  {Campos}}, \bibinfo {author} {\bibfnamefont {D.}~\bibnamefont {Cogollo}},
  \bibinfo {author} {\bibfnamefont {M.}~\bibnamefont {Lindner}}, \bibinfo
  {author} {\bibfnamefont {T.}~\bibnamefont {Melo}}, \bibinfo {author}
  {\bibfnamefont {F.~S.}\ \bibnamefont {Queiroz}}, \ and\ \bibinfo {author}
  {\bibfnamefont {W.}~\bibnamefont {Rodejohann}},\ }\href {\doibase
  10.1007/JHEP08(2017)092} {\bibfield  {journal} {\bibinfo  {journal} {JHEP}\
  }\textbf {\bibinfo {volume} {08}},\ \bibinfo {pages} {092} (\bibinfo {year}
  {2017})},\ \Eprint {http://arxiv.org/abs/1705.05388} {arXiv:1705.05388
  [hep-ph]} \BibitemShut {NoStop}%
\bibitem [{\citenamefont {Camargo}\ \emph {et~al.}(2019)\citenamefont
  {Camargo}, \citenamefont {Dias}, \citenamefont {de~Melo},\ and\ \citenamefont
  {Queiroz}}]{Camargo:2018uzw}%
  \BibitemOpen
  \bibfield  {author} {\bibinfo {author} {\bibfnamefont {D.~A.}\ \bibnamefont
  {Camargo}}, \bibinfo {author} {\bibfnamefont {A.~G.}\ \bibnamefont {Dias}},
  \bibinfo {author} {\bibfnamefont {T.~B.}\ \bibnamefont {de~Melo}}, \ and\
  \bibinfo {author} {\bibfnamefont {F.~S.}\ \bibnamefont {Queiroz}},\ }\href
  {\doibase 10.1007/JHEP04(2019)129} {\bibfield  {journal} {\bibinfo  {journal}
  {JHEP}\ }\textbf {\bibinfo {volume} {04}},\ \bibinfo {pages} {129} (\bibinfo
  {year} {2019})},\ \Eprint {http://arxiv.org/abs/1811.05488} {arXiv:1811.05488
  [hep-ph]} \BibitemShut {NoStop}%
\bibitem [{\citenamefont {Cogollo}\ \emph {et~al.}(2019)\citenamefont
  {Cogollo}, \citenamefont {Matheus}, \citenamefont {de~Melo},\ and\
  \citenamefont {Queiroz}}]{Cogollo:2019mbd}%
  \BibitemOpen
  \bibfield  {author} {\bibinfo {author} {\bibfnamefont {D.}~\bibnamefont
  {Cogollo}}, \bibinfo {author} {\bibfnamefont {R.~D.}\ \bibnamefont
  {Matheus}}, \bibinfo {author} {\bibfnamefont {T.~B.}\ \bibnamefont
  {de~Melo}}, \ and\ \bibinfo {author} {\bibfnamefont {F.~S.}\ \bibnamefont
  {Queiroz}},\ }\href {\doibase 10.1016/j.physletb.2019.134813} {\bibfield
  {journal} {\bibinfo  {journal} {Phys. Lett. B}\ }\textbf {\bibinfo {volume}
  {797}},\ \bibinfo {pages} {134813} (\bibinfo {year} {2019})},\ \Eprint
  {http://arxiv.org/abs/1904.07883} {arXiv:1904.07883 [hep-ph]} \BibitemShut
  {NoStop}%
\bibitem [{\citenamefont {Ko}\ \emph {et~al.}(2014{\natexlab{b}})\citenamefont
  {Ko}, \citenamefont {Omura},\ and\ \citenamefont {Yu}}]{Ko:2013zsa}%
  \BibitemOpen
  \bibfield  {author} {\bibinfo {author} {\bibfnamefont {P.}~\bibnamefont
  {Ko}}, \bibinfo {author} {\bibfnamefont {Y.}~\bibnamefont {Omura}}, \ and\
  \bibinfo {author} {\bibfnamefont {C.}~\bibnamefont {Yu}},\ }\href {\doibase
  10.1007/JHEP01(2014)016} {\bibfield  {journal} {\bibinfo  {journal} {JHEP}\
  }\textbf {\bibinfo {volume} {01}},\ \bibinfo {pages} {016} (\bibinfo {year}
  {2014}{\natexlab{b}})},\ \Eprint {http://arxiv.org/abs/1309.7156}
  {arXiv:1309.7156 [hep-ph]} \BibitemShut {NoStop}%
\bibitem [{\citenamefont {Dine}\ \emph {et~al.}(1981)\citenamefont {Dine},
  \citenamefont {Fischler},\ and\ \citenamefont {Srednicki}}]{Dine:1981rt}%
  \BibitemOpen
  \bibfield  {author} {\bibinfo {author} {\bibfnamefont {M.}~\bibnamefont
  {Dine}}, \bibinfo {author} {\bibfnamefont {W.}~\bibnamefont {Fischler}}, \
  and\ \bibinfo {author} {\bibfnamefont {M.}~\bibnamefont {Srednicki}},\ }\href
  {\doibase 10.1016/0370-2693(81)90590-6} {\bibfield  {journal} {\bibinfo
  {journal} {Phys.Lett.}\ }\textbf {\bibinfo {volume} {B104}},\ \bibinfo
  {pages} {199} (\bibinfo {year} {1981})}\BibitemShut {NoStop}%
\bibitem [{\citenamefont {Zhitnitsky}(1980)}]{Zhitnitsky:1980tq}%
  \BibitemOpen
  \bibfield  {author} {\bibinfo {author} {\bibfnamefont {A.}~\bibnamefont
  {Zhitnitsky}},\ }\href@noop {} {\bibfield  {journal} {\bibinfo  {journal}
  {Sov.J.Nucl.Phys.}\ }\textbf {\bibinfo {volume} {31}},\ \bibinfo {pages}
  {260} (\bibinfo {year} {1980})}\BibitemShut {NoStop}%
\bibitem [{\citenamefont {Quevillon}\ and\ \citenamefont
  {Smith}(2020)}]{Quevillon:2020hmx}%
  \BibitemOpen
  \bibfield  {author} {\bibinfo {author} {\bibfnamefont {J.}~\bibnamefont
  {Quevillon}}\ and\ \bibinfo {author} {\bibfnamefont {C.}~\bibnamefont
  {Smith}},\ }\href {\doibase 10.1103/PhysRevD.102.075031} {\bibfield
  {journal} {\bibinfo  {journal} {Phys. Rev. D}\ }\textbf {\bibinfo {volume}
  {102}},\ \bibinfo {pages} {075031} (\bibinfo {year} {2020})},\ \Eprint
  {http://arxiv.org/abs/2006.06778} {arXiv:2006.06778 [hep-ph]} \BibitemShut
  {NoStop}%
\bibitem [{\citenamefont {Williams}\ \emph {et~al.}(2011)\citenamefont
  {Williams}, \citenamefont {Burgess}, \citenamefont {Maharana},\ and\
  \citenamefont {Quevedo}}]{Williams:2011qb}%
  \BibitemOpen
  \bibfield  {author} {\bibinfo {author} {\bibfnamefont {M.}~\bibnamefont
  {Williams}}, \bibinfo {author} {\bibfnamefont {C.~P.}\ \bibnamefont
  {Burgess}}, \bibinfo {author} {\bibfnamefont {A.}~\bibnamefont {Maharana}}, \
  and\ \bibinfo {author} {\bibfnamefont {F.}~\bibnamefont {Quevedo}},\ }\href
  {\doibase 10.1007/JHEP08(2011)106} {\bibfield  {journal} {\bibinfo  {journal}
  {JHEP}\ }\textbf {\bibinfo {volume} {08}},\ \bibinfo {pages} {106} (\bibinfo
  {year} {2011})},\ \Eprint {http://arxiv.org/abs/1103.4556} {arXiv:1103.4556
  [hep-ph]} \BibitemShut {NoStop}%
\bibitem [{\citenamefont {Sirunyan}\ \emph {et~al.}(2018)\citenamefont
  {Sirunyan} \emph {et~al.}}]{Sirunyan:2018exx}%
  \BibitemOpen
  \bibfield  {author} {\bibinfo {author} {\bibfnamefont {A.~M.}\ \bibnamefont
  {Sirunyan}} \emph {et~al.} (\bibinfo {collaboration} {CMS}),\ }\href
  {\doibase 10.1007/JHEP06(2018)120} {\bibfield  {journal} {\bibinfo  {journal}
  {JHEP}\ }\textbf {\bibinfo {volume} {06}},\ \bibinfo {pages} {120} (\bibinfo
  {year} {2018})},\ \Eprint {http://arxiv.org/abs/1803.06292} {arXiv:1803.06292
  [hep-ex]} \BibitemShut {NoStop}%
\bibitem [{\citenamefont {Aad}\ \emph {et~al.}(2019)\citenamefont {Aad} \emph
  {et~al.}}]{Aad:2019fac}%
  \BibitemOpen
  \bibfield  {author} {\bibinfo {author} {\bibfnamefont {G.}~\bibnamefont
  {Aad}} \emph {et~al.} (\bibinfo {collaboration} {ATLAS}),\ }\href {\doibase
  10.1016/j.physletb.2019.07.016} {\bibfield  {journal} {\bibinfo  {journal}
  {Phys. Lett. B}\ }\textbf {\bibinfo {volume} {796}},\ \bibinfo {pages} {68}
  (\bibinfo {year} {2019})},\ \Eprint {http://arxiv.org/abs/1903.06248}
  {arXiv:1903.06248 [hep-ex]} \BibitemShut {NoStop}%
\bibitem [{\citenamefont {Fileviez~P\'erez}\ \emph {et~al.}(2019)\citenamefont
  {Fileviez~P\'erez}, \citenamefont {Murgui},\ and\ \citenamefont
  {Plascencia}}]{FileviezPerez:2019cyn}%
  \BibitemOpen
  \bibfield  {author} {\bibinfo {author} {\bibfnamefont {P.}~\bibnamefont
  {Fileviez~P\'erez}}, \bibinfo {author} {\bibfnamefont {C.}~\bibnamefont
  {Murgui}}, \ and\ \bibinfo {author} {\bibfnamefont {A.~D.}\ \bibnamefont
  {Plascencia}},\ }\href {\doibase 10.1103/PhysRevD.100.035041} {\bibfield
  {journal} {\bibinfo  {journal} {Phys. Rev. D}\ }\textbf {\bibinfo {volume}
  {100}},\ \bibinfo {pages} {035041} (\bibinfo {year} {2019})},\ \Eprint
  {http://arxiv.org/abs/1905.06344} {arXiv:1905.06344 [hep-ph]} \BibitemShut
  {NoStop}%
\bibitem [{\citenamefont {Schechter}\ and\ \citenamefont
  {Valle}(1982)}]{Schechter:1981cv}%
  \BibitemOpen
  \bibfield  {author} {\bibinfo {author} {\bibfnamefont {J.}~\bibnamefont
  {Schechter}}\ and\ \bibinfo {author} {\bibfnamefont {J.}~\bibnamefont
  {Valle}},\ }\href {\doibase 10.1103/PhysRevD.25.774} {\bibfield  {journal}
  {\bibinfo  {journal} {Phys.Rev.}\ }\textbf {\bibinfo {volume} {D25}},\
  \bibinfo {pages} {774} (\bibinfo {year} {1982})}\BibitemShut {NoStop}%
\bibitem [{\citenamefont {Grimus}\ and\ \citenamefont
  {Lavoura}(2000)}]{Grimus:2000vj}%
  \BibitemOpen
  \bibfield  {author} {\bibinfo {author} {\bibfnamefont {W.}~\bibnamefont
  {Grimus}}\ and\ \bibinfo {author} {\bibfnamefont {L.}~\bibnamefont
  {Lavoura}},\ }\href {\doibase 10.1088/1126-6708/2000/11/042} {\bibfield
  {journal} {\bibinfo  {journal} {JHEP}\ }\textbf {\bibinfo {volume} {11}},\
  \bibinfo {pages} {042} (\bibinfo {year} {2000})},\ \Eprint
  {http://arxiv.org/abs/hep-ph/0008179} {arXiv:hep-ph/0008179 [hep-ph]}
  \BibitemShut {NoStop}%
\bibitem [{\citenamefont {Hettmansperger}\ \emph {et~al.}(2011)\citenamefont
  {Hettmansperger}, \citenamefont {Lindner},\ and\ \citenamefont
  {Rodejohann}}]{Hettmansperger:2011bt}%
  \BibitemOpen
  \bibfield  {author} {\bibinfo {author} {\bibfnamefont {H.}~\bibnamefont
  {Hettmansperger}}, \bibinfo {author} {\bibfnamefont {M.}~\bibnamefont
  {Lindner}}, \ and\ \bibinfo {author} {\bibfnamefont {W.}~\bibnamefont
  {Rodejohann}},\ }\href {\doibase 10.1007/JHEP04(2011)123} {\bibfield
  {journal} {\bibinfo  {journal} {JHEP}\ }\textbf {\bibinfo {volume} {04}},\
  \bibinfo {pages} {123} (\bibinfo {year} {2011})},\ \Eprint
  {http://arxiv.org/abs/1102.3432} {arXiv:1102.3432 [hep-ph]} \BibitemShut
  {NoStop}%
\bibitem [{\citenamefont {Korner}\ \emph {et~al.}(1993)\citenamefont {Korner},
  \citenamefont {Pilaftsis},\ and\ \citenamefont {Schilcher}}]{Korner:1992zk}%
  \BibitemOpen
  \bibfield  {author} {\bibinfo {author} {\bibfnamefont {J.~G.}\ \bibnamefont
  {Korner}}, \bibinfo {author} {\bibfnamefont {A.}~\bibnamefont {Pilaftsis}}, \
  and\ \bibinfo {author} {\bibfnamefont {K.}~\bibnamefont {Schilcher}},\ }\href
  {\doibase 10.1103/PhysRevD.47.1080} {\bibfield  {journal} {\bibinfo
  {journal} {Phys. Rev. D}\ }\textbf {\bibinfo {volume} {47}},\ \bibinfo
  {pages} {1080} (\bibinfo {year} {1993})},\ \Eprint
  {http://arxiv.org/abs/hep-ph/9301289} {arXiv:hep-ph/9301289} \BibitemShut
  {NoStop}%
\bibitem [{\citenamefont {Aguilar-Saavedra}(2003)}]{AguilarSaavedra:2002kr}%
  \BibitemOpen
  \bibfield  {author} {\bibinfo {author} {\bibfnamefont {J.~A.}\ \bibnamefont
  {Aguilar-Saavedra}},\ }\href {\doibase 10.1103/PhysRevD.69.099901} {\bibfield
   {journal} {\bibinfo  {journal} {Phys. Rev. D}\ }\textbf {\bibinfo {volume}
  {67}},\ \bibinfo {pages} {035003} (\bibinfo {year} {2003})},\ \bibinfo {note}
  {[Erratum: Phys.Rev.D 69, 099901 (2004)]},\ \Eprint
  {http://arxiv.org/abs/hep-ph/0210112} {arXiv:hep-ph/0210112} \BibitemShut
  {NoStop}%
\bibitem [{\citenamefont {Vatsyayan}\ and\ \citenamefont
  {Kundu}(2020)}]{Vatsyayan:2020jan}%
  \BibitemOpen
  \bibfield  {author} {\bibinfo {author} {\bibfnamefont {D.}~\bibnamefont
  {Vatsyayan}}\ and\ \bibinfo {author} {\bibfnamefont {A.}~\bibnamefont
  {Kundu}},\ }\href {\doibase 10.1016/j.nuclphysb.2020.115208} {\bibfield
  {journal} {\bibinfo  {journal} {Nucl. Phys. B}\ }\textbf {\bibinfo {volume}
  {960}},\ \bibinfo {pages} {115208} (\bibinfo {year} {2020})},\ \Eprint
  {http://arxiv.org/abs/2007.02327} {arXiv:2007.02327 [hep-ph]} \BibitemShut
  {NoStop}%
\bibitem [{\citenamefont {Srednicki}(1985)}]{Srednicki:1985xd}%
  \BibitemOpen
  \bibfield  {author} {\bibinfo {author} {\bibfnamefont {M.}~\bibnamefont
  {Srednicki}},\ }\href {\doibase 10.1016/0550-3213(85)90054-9} {\bibfield
  {journal} {\bibinfo  {journal} {Nucl.Phys.}\ }\textbf {\bibinfo {volume}
  {B260}},\ \bibinfo {pages} {689} (\bibinfo {year} {1985})}\BibitemShut
  {NoStop}%
\bibitem [{\citenamefont {Di~Luzio}\ \emph {et~al.}(2020)\citenamefont
  {Di~Luzio}, \citenamefont {Giannotti}, \citenamefont {Nardi},\ and\
  \citenamefont {Visinelli}}]{DiLuzio:2020wdo}%
  \BibitemOpen
  \bibfield  {author} {\bibinfo {author} {\bibfnamefont {L.}~\bibnamefont
  {Di~Luzio}}, \bibinfo {author} {\bibfnamefont {M.}~\bibnamefont {Giannotti}},
  \bibinfo {author} {\bibfnamefont {E.}~\bibnamefont {Nardi}}, \ and\ \bibinfo
  {author} {\bibfnamefont {L.}~\bibnamefont {Visinelli}},\ }\href {\doibase
  10.1016/j.physrep.2020.06.002} {\bibfield  {journal} {\bibinfo  {journal}
  {Phys. Rept.}\ }\textbf {\bibinfo {volume} {870}},\ \bibinfo {pages} {1}
  (\bibinfo {year} {2020})},\ \Eprint {http://arxiv.org/abs/2003.01100}
  {arXiv:2003.01100 [hep-ph]} \BibitemShut {NoStop}%
\bibitem [{\citenamefont {Grilli~di Cortona}\ \emph {et~al.}(2016)\citenamefont
  {Grilli~di Cortona}, \citenamefont {Hardy}, \citenamefont {Pardo~Vega},\ and\
  \citenamefont {Villadoro}}]{diCortona:2015ldu}%
  \BibitemOpen
  \bibfield  {author} {\bibinfo {author} {\bibfnamefont {G.}~\bibnamefont
  {Grilli~di Cortona}}, \bibinfo {author} {\bibfnamefont {E.}~\bibnamefont
  {Hardy}}, \bibinfo {author} {\bibfnamefont {J.}~\bibnamefont {Pardo~Vega}}, \
  and\ \bibinfo {author} {\bibfnamefont {G.}~\bibnamefont {Villadoro}},\ }\href
  {\doibase 10.1007/JHEP01(2016)034} {\bibfield  {journal} {\bibinfo  {journal}
  {JHEP}\ }\textbf {\bibinfo {volume} {1601}},\ \bibinfo {pages} {034}
  (\bibinfo {year} {2016})},\ \Eprint {http://arxiv.org/abs/1511.02867}
  {arXiv:1511.02867 [hep-ph]} \BibitemShut {NoStop}%
\bibitem [{\citenamefont {Borsanyi}\ \emph {et~al.}(2016)\citenamefont
  {Borsanyi} \emph {et~al.}}]{Borsanyi:2016ksw}%
  \BibitemOpen
  \bibfield  {author} {\bibinfo {author} {\bibfnamefont {S.}~\bibnamefont
  {Borsanyi}} \emph {et~al.},\ }\href {\doibase 10.1038/nature20115} {\bibfield
   {journal} {\bibinfo  {journal} {Nature}\ }\textbf {\bibinfo {volume}
  {539}},\ \bibinfo {pages} {69} (\bibinfo {year} {2016})},\ \Eprint
  {http://arxiv.org/abs/1606.07494} {arXiv:1606.07494 [hep-lat]} \BibitemShut
  {NoStop}%
\bibitem [{\citenamefont {Gorghetto}\ and\ \citenamefont
  {Villadoro}(2019)}]{Gorghetto:2018ocs}%
  \BibitemOpen
  \bibfield  {author} {\bibinfo {author} {\bibfnamefont {M.}~\bibnamefont
  {Gorghetto}}\ and\ \bibinfo {author} {\bibfnamefont {G.}~\bibnamefont
  {Villadoro}},\ }\href {\doibase 10.1007/JHEP03(2019)033} {\bibfield
  {journal} {\bibinfo  {journal} {JHEP}\ }\textbf {\bibinfo {volume} {03}},\
  \bibinfo {pages} {033} (\bibinfo {year} {2019})},\ \Eprint
  {http://arxiv.org/abs/1812.01008} {arXiv:1812.01008 [hep-ph]} \BibitemShut
  {NoStop}%
\bibitem [{\citenamefont {Zyla}\ \emph {et~al.}(2020)\citenamefont {Zyla} \emph
  {et~al.}}]{Zyla:2020zbs}%
  \BibitemOpen
  \bibfield  {author} {\bibinfo {author} {\bibfnamefont {P.~A.}\ \bibnamefont
  {Zyla}} \emph {et~al.} (\bibinfo {collaboration} {Particle Data Group}),\
  }\href {\doibase 10.1093/ptep/ptaa104} {\bibfield  {journal} {\bibinfo
  {journal} {PTEP}\ }\textbf {\bibinfo {volume} {2020}},\ \bibinfo {pages}
  {083C01} (\bibinfo {year} {2020})}\BibitemShut {NoStop}%
\bibitem [{\citenamefont {Sikivie}(2021)}]{Sikivie:2020zpn}%
  \BibitemOpen
  \bibfield  {author} {\bibinfo {author} {\bibfnamefont {P.}~\bibnamefont
  {Sikivie}},\ }\href {\doibase 10.1103/RevModPhys.93.015004} {\bibfield
  {journal} {\bibinfo  {journal} {Rev. Mod. Phys.}\ }\textbf {\bibinfo {volume}
  {93}},\ \bibinfo {pages} {015004} (\bibinfo {year} {2021})},\ \Eprint
  {http://arxiv.org/abs/2003.02206} {arXiv:2003.02206 [hep-ph]} \BibitemShut
  {NoStop}%
\bibitem [{\citenamefont {Akrami}\ \emph {et~al.}(2020)\citenamefont {Akrami}
  \emph {et~al.}}]{Akrami:2018odb}%
  \BibitemOpen
  \bibfield  {author} {\bibinfo {author} {\bibfnamefont {Y.}~\bibnamefont
  {Akrami}} \emph {et~al.} (\bibinfo {collaboration} {Planck}),\ }\href
  {\doibase 10.1051/0004-6361/201833887} {\bibfield  {journal} {\bibinfo
  {journal} {Astron. Astrophys.}\ }\textbf {\bibinfo {volume} {641}},\ \bibinfo
  {pages} {A10} (\bibinfo {year} {2020})},\ \Eprint
  {http://arxiv.org/abs/1807.06211} {arXiv:1807.06211 [astro-ph.CO]}
  \BibitemShut {NoStop}%
\bibitem [{\citenamefont {Zioutas}\ \emph {et~al.}(2005)\citenamefont {Zioutas}
  \emph {et~al.}}]{Zioutas:2004hi}%
  \BibitemOpen
  \bibfield  {author} {\bibinfo {author} {\bibfnamefont {K.}~\bibnamefont
  {Zioutas}} \emph {et~al.} (\bibinfo {collaboration} {CAST}),\ }\href
  {\doibase 10.1103/PhysRevLett.94.121301} {\bibfield  {journal} {\bibinfo
  {journal} {Phys. Rev. Lett.}\ }\textbf {\bibinfo {volume} {94}},\ \bibinfo
  {pages} {121301} (\bibinfo {year} {2005})},\ \Eprint
  {http://arxiv.org/abs/hep-ex/0411033} {arXiv:hep-ex/0411033} \BibitemShut
  {NoStop}%
\bibitem [{\citenamefont {Andriamonje}\ \emph {et~al.}(2007)\citenamefont
  {Andriamonje} \emph {et~al.}}]{Andriamonje:2007ew}%
  \BibitemOpen
  \bibfield  {author} {\bibinfo {author} {\bibfnamefont {S.}~\bibnamefont
  {Andriamonje}} \emph {et~al.} (\bibinfo {collaboration} {CAST}),\ }\href
  {\doibase 10.1088/1475-7516/2007/04/010} {\bibfield  {journal} {\bibinfo
  {journal} {JCAP}\ }\textbf {\bibinfo {volume} {04}},\ \bibinfo {pages} {010}
  (\bibinfo {year} {2007})},\ \Eprint {http://arxiv.org/abs/hep-ex/0702006}
  {arXiv:hep-ex/0702006} \BibitemShut {NoStop}%
\bibitem [{\citenamefont {Anastassopoulos}\ \emph {et~al.}(2017)\citenamefont
  {Anastassopoulos} \emph {et~al.}}]{Anastassopoulos:2017ftl}%
  \BibitemOpen
  \bibfield  {author} {\bibinfo {author} {\bibfnamefont {V.}~\bibnamefont
  {Anastassopoulos}} \emph {et~al.} (\bibinfo {collaboration} {CAST}),\ }\href
  {\doibase 10.1038/nphys4109} {\bibfield  {journal} {\bibinfo  {journal}
  {Nature Phys.}\ }\textbf {\bibinfo {volume} {13}},\ \bibinfo {pages} {584}
  (\bibinfo {year} {2017})},\ \Eprint {http://arxiv.org/abs/1705.02290}
  {arXiv:1705.02290 [hep-ex]} \BibitemShut {NoStop}%
\bibitem [{\citenamefont {Duffy}\ \emph {et~al.}(2006)\citenamefont {Duffy}
  \emph {et~al.}}]{Duffy:2006aa}%
  \BibitemOpen
  \bibfield  {author} {\bibinfo {author} {\bibfnamefont {L.~D.}\ \bibnamefont
  {Duffy}} \emph {et~al.} (\bibinfo {collaboration} {ADMX}),\ }\href {\doibase
  10.1103/PhysRevD.74.012006} {\bibfield  {journal} {\bibinfo  {journal}
  {Phys.Rev.}\ }\textbf {\bibinfo {volume} {D74}},\ \bibinfo {pages} {012006}
  (\bibinfo {year} {2006})}\BibitemShut {NoStop}%
\bibitem [{\citenamefont {Asztalos}\ \emph {et~al.}(2010)\citenamefont
  {Asztalos} \emph {et~al.}}]{Asztalos:2009yp}%
  \BibitemOpen
  \bibfield  {author} {\bibinfo {author} {\bibfnamefont {S.}~\bibnamefont
  {Asztalos}} \emph {et~al.} (\bibinfo {collaboration} {ADMX}),\ }\href
  {\doibase 10.1103/PhysRevLett.104.041301} {\bibfield  {journal} {\bibinfo
  {journal} {Phys.Rev.Lett.}\ }\textbf {\bibinfo {volume} {104}},\ \bibinfo
  {pages} {041301} (\bibinfo {year} {2010})},\ \Eprint
  {http://arxiv.org/abs/0910.5914} {arXiv:0910.5914 [astro-ph.CO]} \BibitemShut
  {NoStop}%
\bibitem [{\citenamefont {Asztalos}\ \emph {et~al.}(2011)\citenamefont
  {Asztalos} \emph {et~al.}}]{Asztalos:2011bm}%
  \BibitemOpen
  \bibfield  {author} {\bibinfo {author} {\bibfnamefont {S.}~\bibnamefont
  {Asztalos}} \emph {et~al.} (\bibinfo {collaboration} {ADMX}),\ }\href
  {\doibase 10.1016/j.nima.2011.07.019} {\bibfield  {journal} {\bibinfo
  {journal} {Nucl.Instrum.Meth.}\ }\textbf {\bibinfo {volume} {A656}},\
  \bibinfo {pages} {39} (\bibinfo {year} {2011})},\ \Eprint
  {http://arxiv.org/abs/1105.4203} {arXiv:1105.4203 [physics.ins-det]}
  \BibitemShut {NoStop}%
\bibitem [{\citenamefont {Stern}(2016)}]{Stern:2016bbw}%
  \BibitemOpen
  \bibfield  {author} {\bibinfo {author} {\bibfnamefont {I.}~\bibnamefont
  {Stern}}\ }(\bibinfo {year} {2016})\ p.\ \bibinfo {pages} {198},\ \Eprint
  {http://arxiv.org/abs/1612.08296} {arXiv:1612.08296 [physics.ins-det]}
  \BibitemShut {NoStop}%
\bibitem [{\citenamefont {Braine}\ \emph {et~al.}(2020)\citenamefont {Braine}
  \emph {et~al.}}]{Braine:2019fqb}%
  \BibitemOpen
  \bibfield  {author} {\bibinfo {author} {\bibfnamefont {T.}~\bibnamefont
  {Braine}} \emph {et~al.} (\bibinfo {collaboration} {ADMX}),\ }\href {\doibase
  10.1103/PhysRevLett.124.101303} {\bibfield  {journal} {\bibinfo  {journal}
  {Phys.Rev.Lett.}\ }\textbf {\bibinfo {volume} {124}},\ \bibinfo {pages}
  {101303} (\bibinfo {year} {2020})},\ \Eprint
  {http://arxiv.org/abs/1910.08638} {arXiv:1910.08638 [hep-ex]} \BibitemShut
  {NoStop}%
\bibitem [{\citenamefont {Ayala}\ \emph {et~al.}(2014)\citenamefont {Ayala},
  \citenamefont {Dom\'\i{}nguez}, \citenamefont {Giannotti}, \citenamefont
  {Mirizzi},\ and\ \citenamefont {Straniero}}]{Ayala:2014pea}%
  \BibitemOpen
  \bibfield  {author} {\bibinfo {author} {\bibfnamefont {A.}~\bibnamefont
  {Ayala}}, \bibinfo {author} {\bibfnamefont {I.}~\bibnamefont
  {Dom\'\i{}nguez}}, \bibinfo {author} {\bibfnamefont {M.}~\bibnamefont
  {Giannotti}}, \bibinfo {author} {\bibfnamefont {A.}~\bibnamefont {Mirizzi}},
  \ and\ \bibinfo {author} {\bibfnamefont {O.}~\bibnamefont {Straniero}},\
  }\href {\doibase 10.1103/PhysRevLett.113.191302} {\bibfield  {journal}
  {\bibinfo  {journal} {Phys. Rev. Lett.}\ }\textbf {\bibinfo {volume} {113}},\
  \bibinfo {pages} {191302} (\bibinfo {year} {2014})},\ \Eprint
  {http://arxiv.org/abs/1406.6053} {arXiv:1406.6053 [astro-ph.SR]} \BibitemShut
  {NoStop}%
\bibitem [{\citenamefont {Straniero}\ \emph {et~al.}(2015)\citenamefont
  {Straniero}, \citenamefont {Ayala}, \citenamefont {Giannotti}, \citenamefont
  {Mirizzi},\ and\ \citenamefont {Dominguez}}]{Straniero:2015nvc}%
  \BibitemOpen
  \bibfield  {author} {\bibinfo {author} {\bibfnamefont {O.}~\bibnamefont
  {Straniero}}, \bibinfo {author} {\bibfnamefont {A.}~\bibnamefont {Ayala}},
  \bibinfo {author} {\bibfnamefont {M.}~\bibnamefont {Giannotti}}, \bibinfo
  {author} {\bibfnamefont {A.}~\bibnamefont {Mirizzi}}, \ and\ \bibinfo
  {author} {\bibfnamefont {I.}~\bibnamefont {Dominguez}},\ }in\ \href {\doibase
  10.3204/DESY-PROC-2015-02/straniero_oscar} {\emph {\bibinfo {booktitle}
  {{11th Patras Workshop on Axions, WIMPs and WISPs}}}}\ (\bibinfo  {publisher}
  {Deutsches Elektronen-Synchrotron Hamburg},\ \bibinfo {address} {Hamburg,
  Germany},\ \bibinfo {year} {2015})\BibitemShut {NoStop}%
\bibitem [{\citenamefont {Hannestad}\ \emph {et~al.}(2010)\citenamefont
  {Hannestad}, \citenamefont {Mirizzi}, \citenamefont {Raffelt},\ and\
  \citenamefont {Wong}}]{Hannestad:2010yi}%
  \BibitemOpen
  \bibfield  {author} {\bibinfo {author} {\bibfnamefont {S.}~\bibnamefont
  {Hannestad}}, \bibinfo {author} {\bibfnamefont {A.}~\bibnamefont {Mirizzi}},
  \bibinfo {author} {\bibfnamefont {G.~G.}\ \bibnamefont {Raffelt}}, \ and\
  \bibinfo {author} {\bibfnamefont {Y.~Y.~Y.}\ \bibnamefont {Wong}},\ }\href
  {\doibase 10.1088/1475-7516/2010/08/001} {\bibfield  {journal} {\bibinfo
  {journal} {JCAP}\ }\textbf {\bibinfo {volume} {08}},\ \bibinfo {pages} {001}
  (\bibinfo {year} {2010})},\ \Eprint {http://arxiv.org/abs/1004.0695}
  {arXiv:1004.0695 [astro-ph.CO]} \BibitemShut {NoStop}%
\bibitem [{\citenamefont {Archidiacono}\ \emph {et~al.}(2013)\citenamefont
  {Archidiacono}, \citenamefont {Hannestad}, \citenamefont {Mirizzi},
  \citenamefont {Raffelt},\ and\ \citenamefont {Wong}}]{Archidiacono:2013cha}%
  \BibitemOpen
  \bibfield  {author} {\bibinfo {author} {\bibfnamefont {M.}~\bibnamefont
  {Archidiacono}}, \bibinfo {author} {\bibfnamefont {S.}~\bibnamefont
  {Hannestad}}, \bibinfo {author} {\bibfnamefont {A.}~\bibnamefont {Mirizzi}},
  \bibinfo {author} {\bibfnamefont {G.}~\bibnamefont {Raffelt}}, \ and\
  \bibinfo {author} {\bibfnamefont {Y.~Y.~Y.}\ \bibnamefont {Wong}},\ }\href
  {\doibase 10.1088/1475-7516/2013/10/020} {\bibfield  {journal} {\bibinfo
  {journal} {JCAP}\ }\textbf {\bibinfo {volume} {10}},\ \bibinfo {pages} {020}
  (\bibinfo {year} {2013})},\ \Eprint {http://arxiv.org/abs/1307.0615}
  {arXiv:1307.0615 [astro-ph.CO]} \BibitemShut {NoStop}%
\bibitem [{\citenamefont {Di~Valentino}\ \emph {et~al.}(2016)\citenamefont
  {Di~Valentino}, \citenamefont {Giusarma}, \citenamefont {Lattanzi},
  \citenamefont {Mena}, \citenamefont {Melchiorri},\ and\ \citenamefont
  {Silk}}]{DiValentino:2015wba}%
  \BibitemOpen
  \bibfield  {author} {\bibinfo {author} {\bibfnamefont {E.}~\bibnamefont
  {Di~Valentino}}, \bibinfo {author} {\bibfnamefont {E.}~\bibnamefont
  {Giusarma}}, \bibinfo {author} {\bibfnamefont {M.}~\bibnamefont {Lattanzi}},
  \bibinfo {author} {\bibfnamefont {O.}~\bibnamefont {Mena}}, \bibinfo {author}
  {\bibfnamefont {A.}~\bibnamefont {Melchiorri}}, \ and\ \bibinfo {author}
  {\bibfnamefont {J.}~\bibnamefont {Silk}},\ }\href {\doibase
  10.1016/j.physletb.2015.11.025} {\bibfield  {journal} {\bibinfo  {journal}
  {Phys. Lett. B}\ }\textbf {\bibinfo {volume} {752}},\ \bibinfo {pages} {182}
  (\bibinfo {year} {2016})},\ \Eprint {http://arxiv.org/abs/1507.08665}
  {arXiv:1507.08665 [astro-ph.CO]} \BibitemShut {NoStop}%
\bibitem [{\citenamefont {Kahn}\ \emph {et~al.}(2016)\citenamefont {Kahn},
  \citenamefont {Safdi},\ and\ \citenamefont {Thaler}}]{Kahn:2016aff}%
  \BibitemOpen
  \bibfield  {author} {\bibinfo {author} {\bibfnamefont {Y.}~\bibnamefont
  {Kahn}}, \bibinfo {author} {\bibfnamefont {B.~R.}\ \bibnamefont {Safdi}}, \
  and\ \bibinfo {author} {\bibfnamefont {J.}~\bibnamefont {Thaler}},\ }\href
  {\doibase 10.1103/PhysRevLett.117.141801} {\bibfield  {journal} {\bibinfo
  {journal} {Phys. Rev. Lett.}\ }\textbf {\bibinfo {volume} {117}},\ \bibinfo
  {pages} {141801} (\bibinfo {year} {2016})},\ \Eprint
  {http://arxiv.org/abs/1602.01086} {arXiv:1602.01086 [hep-ph]} \BibitemShut
  {NoStop}%
\bibitem [{\citenamefont {Ouellet}\ \emph {et~al.}(2019)\citenamefont {Ouellet}
  \emph {et~al.}}]{Ouellet:2018beu}%
  \BibitemOpen
  \bibfield  {author} {\bibinfo {author} {\bibfnamefont {J.~L.}\ \bibnamefont
  {Ouellet}} \emph {et~al.},\ }\href {\doibase 10.1103/PhysRevLett.122.121802}
  {\bibfield  {journal} {\bibinfo  {journal} {Phys. Rev. Lett.}\ }\textbf
  {\bibinfo {volume} {122}},\ \bibinfo {pages} {121802} (\bibinfo {year}
  {2019})},\ \Eprint {http://arxiv.org/abs/1810.12257} {arXiv:1810.12257
  [hep-ex]} \BibitemShut {NoStop}%
\bibitem [{\citenamefont {Caldwell}\ \emph {et~al.}(2017)\citenamefont
  {Caldwell}, \citenamefont {Dvali}, \citenamefont {Majorovits}, \citenamefont
  {Millar}, \citenamefont {Raffelt}, \citenamefont {Redondo}, \citenamefont
  {Reimann}, \citenamefont {Simon},\ and\ \citenamefont
  {Steffen}}]{TheMADMAXWorkingGroup:2016hpc}%
  \BibitemOpen
  \bibfield  {author} {\bibinfo {author} {\bibfnamefont {A.}~\bibnamefont
  {Caldwell}}, \bibinfo {author} {\bibfnamefont {G.}~\bibnamefont {Dvali}},
  \bibinfo {author} {\bibfnamefont {B.}~\bibnamefont {Majorovits}}, \bibinfo
  {author} {\bibfnamefont {A.}~\bibnamefont {Millar}}, \bibinfo {author}
  {\bibfnamefont {G.}~\bibnamefont {Raffelt}}, \bibinfo {author} {\bibfnamefont
  {J.}~\bibnamefont {Redondo}}, \bibinfo {author} {\bibfnamefont
  {O.}~\bibnamefont {Reimann}}, \bibinfo {author} {\bibfnamefont
  {F.}~\bibnamefont {Simon}}, \ and\ \bibinfo {author} {\bibfnamefont
  {F.}~\bibnamefont {Steffen}} (\bibinfo {collaboration} {MADMAX Working
  Group}),\ }\href {\doibase 10.1103/PhysRevLett.118.091801} {\bibfield
  {journal} {\bibinfo  {journal} {Phys. Rev. Lett.}\ }\textbf {\bibinfo
  {volume} {118}},\ \bibinfo {pages} {091801} (\bibinfo {year} {2017})},\
  \Eprint {http://arxiv.org/abs/1611.05865} {arXiv:1611.05865
  [physics.ins-det]} \BibitemShut {NoStop}%
\bibitem [{\citenamefont {Armengaud}\ \emph {et~al.}(2019)\citenamefont
  {Armengaud} \emph {et~al.}}]{Armengaud:2019uso}%
  \BibitemOpen
  \bibfield  {author} {\bibinfo {author} {\bibfnamefont {E.}~\bibnamefont
  {Armengaud}} \emph {et~al.} (\bibinfo {collaboration} {IAXO}),\ }\href
  {\doibase 10.1088/1475-7516/2019/06/047} {\bibfield  {journal} {\bibinfo
  {journal} {JCAP}\ }\textbf {\bibinfo {volume} {1906}},\ \bibinfo {pages}
  {047} (\bibinfo {year} {2019})},\ \Eprint {http://arxiv.org/abs/1904.09155}
  {arXiv:1904.09155 [hep-ph]} \BibitemShut {NoStop}%
\bibitem [{\citenamefont {Ortiz}\ \emph {et~al.}(2020)\citenamefont {Ortiz}
  \emph {et~al.}}]{Hartman:2020bgj}%
  \BibitemOpen
  \bibfield  {author} {\bibinfo {author} {\bibfnamefont {M.~D.}\ \bibnamefont
  {Ortiz}} \emph {et~al.},\ }\href@noop {} {\  (\bibinfo {year} {2020})},\
  \Eprint {http://arxiv.org/abs/2009.14294} {arXiv:2009.14294 [physics.optics]}
  \BibitemShut {NoStop}%
\bibitem [{\citenamefont {B\"ahre}\ \emph {et~al.}(2013)\citenamefont {B\"ahre}
  \emph {et~al.}}]{Bahre:2013ywa}%
  \BibitemOpen
  \bibfield  {author} {\bibinfo {author} {\bibfnamefont {R.}~\bibnamefont
  {B\"ahre}} \emph {et~al.},\ }\href {\doibase 10.1088/1748-0221/8/09/T09001}
  {\bibfield  {journal} {\bibinfo  {journal} {JINST}\ }\textbf {\bibinfo
  {volume} {8}},\ \bibinfo {pages} {T09001} (\bibinfo {year} {2013})},\ \Eprint
  {http://arxiv.org/abs/1302.5647} {arXiv:1302.5647 [physics.ins-det]}
  \BibitemShut {NoStop}%
\bibitem [{\citenamefont {Bastidon}(2015)}]{Bastidon:2015efa}%
  \BibitemOpen
  \bibfield  {author} {\bibinfo {author} {\bibfnamefont {N.}~\bibnamefont
  {Bastidon}} (\bibinfo {collaboration} {ALPS II}),\ }in\ \href {\doibase
  10.3204/DESY-PROC-2015-02/bastidon_noemie_talk} {\emph {\bibinfo {booktitle}
  {{11th Patras Workshop on Axions, WIMPs and WISPs}}}}\ (\bibinfo  {publisher}
  {Deutsches Elektronen-Synchrotron Hamburg},\ \bibinfo {address} {Hamburg,
  Germany},\ \bibinfo {year} {2015})\ \Eprint {http://arxiv.org/abs/1509.02070}
  {arXiv:1509.02070 [physics.ins-det]} \BibitemShut {NoStop}%
\bibitem [{\citenamefont {Vagnozzi}\ \emph {et~al.}(2017)\citenamefont
  {Vagnozzi}, \citenamefont {Giusarma}, \citenamefont {Mena}, \citenamefont
  {Freese}, \citenamefont {Gerbino}, \citenamefont {Ho},\ and\ \citenamefont
  {Lattanzi}}]{Vagnozzi:2017ovm}%
  \BibitemOpen
  \bibfield  {author} {\bibinfo {author} {\bibfnamefont {S.}~\bibnamefont
  {Vagnozzi}}, \bibinfo {author} {\bibfnamefont {E.}~\bibnamefont {Giusarma}},
  \bibinfo {author} {\bibfnamefont {O.}~\bibnamefont {Mena}}, \bibinfo {author}
  {\bibfnamefont {K.}~\bibnamefont {Freese}}, \bibinfo {author} {\bibfnamefont
  {M.}~\bibnamefont {Gerbino}}, \bibinfo {author} {\bibfnamefont
  {S.}~\bibnamefont {Ho}}, \ and\ \bibinfo {author} {\bibfnamefont
  {M.}~\bibnamefont {Lattanzi}},\ }\href {\doibase 10.1103/PhysRevD.96.123503}
  {\bibfield  {journal} {\bibinfo  {journal} {Phys. Rev. D}\ }\textbf {\bibinfo
  {volume} {96}},\ \bibinfo {pages} {123503} (\bibinfo {year} {2017})},\
  \Eprint {http://arxiv.org/abs/1701.08172} {arXiv:1701.08172 [astro-ph.CO]}
  \BibitemShut {NoStop}%
\bibitem [{\citenamefont {Aker}\ \emph {et~al.}(2019)\citenamefont {Aker} \emph
  {et~al.}}]{Aker:2019uuj}%
  \BibitemOpen
  \bibfield  {author} {\bibinfo {author} {\bibfnamefont {M.}~\bibnamefont
  {Aker}} \emph {et~al.} (\bibinfo {collaboration} {KATRIN}),\ }\href {\doibase
  10.1103/PhysRevLett.123.221802} {\bibfield  {journal} {\bibinfo  {journal}
  {Phys. Rev. Lett.}\ }\textbf {\bibinfo {volume} {123}},\ \bibinfo {pages}
  {221802} (\bibinfo {year} {2019})},\ \Eprint
  {http://arxiv.org/abs/1909.06048} {arXiv:1909.06048 [hep-ex]} \BibitemShut
  {NoStop}%
\bibitem [{\citenamefont {Ema}\ \emph {et~al.}(2017)\citenamefont {Ema},
  \citenamefont {Hamaguchi}, \citenamefont {Moroi},\ and\ \citenamefont
  {Nakayama}}]{Ema:2016ops}%
  \BibitemOpen
  \bibfield  {author} {\bibinfo {author} {\bibfnamefont {Y.}~\bibnamefont
  {Ema}}, \bibinfo {author} {\bibfnamefont {K.}~\bibnamefont {Hamaguchi}},
  \bibinfo {author} {\bibfnamefont {T.}~\bibnamefont {Moroi}}, \ and\ \bibinfo
  {author} {\bibfnamefont {K.}~\bibnamefont {Nakayama}},\ }\href {\doibase
  10.1007/JHEP01(2017)096} {\bibfield  {journal} {\bibinfo  {journal} {JHEP}\
  }\textbf {\bibinfo {volume} {01}},\ \bibinfo {pages} {096} (\bibinfo {year}
  {2017})},\ \Eprint {http://arxiv.org/abs/1612.05492} {arXiv:1612.05492
  [hep-ph]} \BibitemShut {NoStop}%
\bibitem [{\citenamefont {Bj\"orkeroth}\ \emph {et~al.}(2018)\citenamefont
  {Bj\"orkeroth}, \citenamefont {Chun},\ and\ \citenamefont
  {King}}]{Bjorkeroth:2018dzu}%
  \BibitemOpen
  \bibfield  {author} {\bibinfo {author} {\bibfnamefont {F.}~\bibnamefont
  {Bj\"orkeroth}}, \bibinfo {author} {\bibfnamefont {E.~J.}\ \bibnamefont
  {Chun}}, \ and\ \bibinfo {author} {\bibfnamefont {S.~F.}\ \bibnamefont
  {King}},\ }\href {\doibase 10.1007/JHEP08(2018)117} {\bibfield  {journal}
  {\bibinfo  {journal} {JHEP}\ }\textbf {\bibinfo {volume} {08}},\ \bibinfo
  {pages} {117} (\bibinfo {year} {2018})},\ \Eprint
  {http://arxiv.org/abs/1806.00660} {arXiv:1806.00660 [hep-ph]} \BibitemShut
  {NoStop}%
\bibitem [{\citenamefont {Adler}\ \emph {et~al.}(2008)\citenamefont {Adler}
  \emph {et~al.}}]{Adler:2008zza}%
  \BibitemOpen
  \bibfield  {author} {\bibinfo {author} {\bibfnamefont {S.}~\bibnamefont
  {Adler}} \emph {et~al.} (\bibinfo {collaboration} {E949, E787}),\ }\href
  {\doibase 10.1103/PhysRevD.77.052003} {\bibfield  {journal} {\bibinfo
  {journal} {Phys. Rev. D}\ }\textbf {\bibinfo {volume} {77}},\ \bibinfo
  {pages} {052003} (\bibinfo {year} {2008})},\ \Eprint
  {http://arxiv.org/abs/0709.1000} {arXiv:0709.1000 [hep-ex]} \BibitemShut
  {NoStop}%
\end{thebibliography}%

\end{document}